\documentclass[a4paper,12pt]{article}
\usepackage{amsfonts}
\usepackage{amsmath}
\usepackage{amssymb}
\usepackage{latexsym}
\usepackage{color}
\usepackage{colordvi}
\usepackage{epsfig}
\usepackage{array}
\usepackage{pifont}
\usepackage{cite}

\usepackage{geometry}
\usepackage{amsxtra}
\usepackage{axodraw}
\usepackage{amsfonts}
\usepackage{amsmath}
\usepackage{amssymb}
\usepackage{latexsym}
\usepackage{color}
\usepackage{colordvi}
\usepackage{epsfig}
\usepackage{array}
\usepackage{pifont}
\usepackage{axodraw}

\newcommand{\pslash}{p \!\!\!/}

\newcommand{\Aslash}{A \!\!\!/}

\newcommand{\beq}{\begin{eqnarray}}
\newcommand{\eeq}{\end{eqnarray}}

\newcommand{\MSB}{\overline{\rm MS}}

\newcommand{\mMSB}{\mathrm{\overline{\rm MS}}}
\newcommand{\mWMSB}{\mathrm{W\overline{\rm MS}}}

\newcommand{\be}{\begin{equation}}
\newcommand{\ee}{\end{equation}}
\newcommand{\lwrsim}{\raise0.3ex\hbox{$<$\kern-0.75em\raise-1.1ex\hbox{$\sim$}}}

\newcommand{\bea}{\begin{eqnarray}}
\newcommand{\eea}{\end{eqnarray}}
\newcommand{\no}{\nonumber \\}

\def\C2#1#2{({\cal C}_2)_{#1}^{#2}}

\def\eq#1{Eq.~(\ref{#1})}

\def\VEV#1{\langle #1 \rangle}

\def\npb#1#2#3{Nucl.\ Phys.\ {\bf B#1} (#2) #3}
\def\plb#1#2#3{Phys.\ Lett.\ {\bf B#1} (#2) #3}

\def\fig#1{Fig. \ref{#1}}

\newcommand{\tr}{\text{Tr}}
\newcommand{\re}{\mathbb{R}\text{e}}

\numberwithin{equation}{section} 
\title{Renormalisation of quark propagators  from
twisted-mass lattice QCD at $N_f$=2} 

\author{B. Blossier$^a$, Ph.~Boucaud$^a$, M.
Brinet$^b$, F. De Soto$^c$, Z. Liu$^{d,e}$, V.~Morenas$^f$\\ O.
P\`ene$^a$, K. Petrov$^a$, J.~Rodr\'iguez-Quintero$^g$   }

\date{}

\begin{document}

\maketitle
\begin{figure}[h]
  \begin{center}
    \includegraphics[width=50mm]{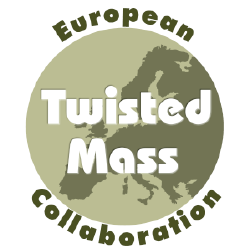}
  \end{center}
\end{figure}

\begin{center}
$^a$ Laboratoire de Physique Th\'eorique\footnote{Unit\'e Mixte de Recherche 8627 du Centre National de
la Recherche Scientifique},\\
CNRS et Universit\'e  Paris-Sud XI, B\^atiment 210, 91405 Orsay Cedex,
France\\
$^b$ Laboratoire de Physique Subatomique et de Cosmologie, CNRS/IN2P3/UJF, \\
53, avenue des Martyrs, 38026 Grenoble, France\\
$^c$ Dpto. Sistemas F\'isicos, Qu\'imicos y Naturales, 
Univ. Pablo de Olavide, 41013 Sevilla, Spain\\
$^d$ DAMTP, University of Cambridge, 
Wilberforce Road, Cambridge CB3 0WA, United Kingdom\\
$^e$ Institute of High Energy Physics,  
Chinese Academy of Science, Beijing 100049, China \\
$^f$ Laboratoire de Physique Corpusculaire, Universit\'e Blaise Pascal, CNRS/IN2P3 \\
63177 Aubi\`ere Cedex, France\\
$^f$ Dpto. F\'isica Aplicada, Fac. Ciencias Experimentales, 
Univ. de Huelva, 21071 Huelva, Spain
\end{center}

\newpage
\begin{abstract}

 We present  results concerning the non-perturbative evaluation of
the renormalisation constant for the quark field, $Z_q$, from lattice 
simulations with twisted mass quarks and three values of the lattice spacing. 
We use the  RI'-MOM scheme.   $Z_q$ has very large lattice spacing artefacts ; 
it  is  considered here as a test bed to elaborate accurate
methods  which will be used for other renormalisation constants.  
We recall and develop the non-perturbative
correction methods  and propose tools to test the quality of the correction.
These tests are also applied to the perturbative  correction method. We check
 that the lattice spacing artefacts scale indeed as $a^2p^2$.

 We then study  the running of $Z_q$ with  particular attention to the
non-perturbative effects, presumably dominated by the dimension-two gluon 
condensate $\VEV{A^2}$ in Landau gauge. We show indeed that this effect is
present, and not small. We check its scaling  in physical units confirming that
it is a continuum effect.  It gives a $\sim 4 \%$  contribution at 2 GeV. 

Different variants are used in order to test the reliability of our result
and estimate the systematic uncertainties. 

Finally combining all our results and using the known Wilson
coefficient of $\VEV{A^2}$ we find  $g^2(\mu^2) \VEV{A^2}_{\mu^2\; CM} =
 2.01(11)\left(^{+0.61}_{- 0.73}\right) \;\mathrm {GeV}^2$ at $\mu=10\,
\mathrm{GeV}$, in fair agreement within uncertainties 
with the value indepently extracted from the strong coupling constant.  We
convert the non-perturbative part of $Z_q$ from RI'-MOM to
$\mMSB$.  Our result for the quark field renormalisation constant in the
$\overline{\mathrm {MS}}$ scheme is $Z_q^{\overline{\mathrm {MS}}\,\mathrm{pert}}((2\,{\mathrm {GeV}})^2,
g^2_{\mathrm {bare}}) = 0.750(3)(7) - 0.313(20) 
\,(g^2_{\mathrm {bare}}-1.5)$ for the perturbative contribution and 
$Z_q^{\overline{\mathrm {MS}}\,\mathrm{non-perturbative}}((2\,{\mathrm {GeV}})^2,
g^2_{\mathrm {bare}}) = 0.781(6)(21) - 0.313(20) \,(g^2_{\mathrm {bare}}-1.5)$
when the non-perturbative contribution is included.
\end{abstract}

\begin{flushleft}
DAMTP-2010-88\\
LPT-Orsay 10-81\\
UHU-FT/10-39 \\
LPSC-10122 \\
PCCF-1007

\end{flushleft}

\tableofcontents



\newcommand{\ewxy}[2]{\setlength{\epsfxsize}{#2}\epsfbox[-40 60 640
590]{#1}}
\newcommand{\ghostSD}{\begin{picture}(150,25)(0,0)
\SetWidth{1.2}
\DashArrowLine(12.5,0)(37.5,0){5}
\DashArrowLine(37.5,0)(75,0){5}
\DashLine(75,0)(112.5,0){5}
\DashArrowLine(112.5,0)(137.5,0){5}
\SetWidth{1}
\Vertex(112.5,0){2}
\GlueArc(75,0)(37.5,0,90){-4}{6}
\GlueArc(75,0)(37.5,90,180){-4}{6}
\CCirc(75,0){5}{Black}{Yellow}
\CCirc(75,37.5){5}{Black}{Yellow}
\CCirc(37.5,0){5}{Black}{Yellow}
\Text(20,-10)[l]{a,k}
\Text(50,15)[l]{d,$\nu$}
\Text(100,-10)[l]{e}
\Text(100,15)[r]{f,$\mu$}
\Text(50,-10)[l]{c,q}
\Text(120,-10)[l]{b,k}
\Text(75,48)[c]{q-k}
\end{picture}}
\newcommand{\ghostDr}{\begin{picture}(100,25)(0,0)
\SetWidth{1.2}
\DashArrowLine(12.5,0)(50,0){5}
\DashArrowLine(50,0)(87.5,0){5}
\CCirc(50,0){5}{Black}{Yellow}
\Text(12.5,-10)[l]{a}
\Text(87.5,-10)[r]{b}
\Text(50,-10)[c]{k}
\end{picture}}
\newcommand{\ghostBr}{\begin{picture}(100,25)(0,0)
\SetWidth{1.2}
\DashArrowLine(12.5,0)(87.5,0){5}
\Text(12.5,-10)[l]{a}
\Text(87.5,-10)[r]{b}
\Text(50,-10)[c]{k}
\end{picture}}
\newcommand{\quark}{\begin{picture}(100,25)(0,0)
\SetWidth{1.2}
\ArrowLine(12.5,0)(50,0)
\ArrowLine(50,0)(92.5,0)
\ArrowLine(92.5,0)(125,0)
\Gluon(50,3)(50,25){4}{3}
\Gluon(87.5,3)(87.5,25){4}{3}
\CCirc(50,0){3}{Black}{Yellow}
\CCirc(87.5,0){3}{Black}{Yellow}
\CCirc(50,25){5}{Black}{Red}
\CCirc(87.5,25){5}{Black}{Red}
\Text(25,+10)[l]{p}
\Text(70,+10)[l]{p}
\Text(100,+10)[l]{p}
\end{picture}}
\newcommand{\vertex}{\begin{picture}(100,25)(0,0)
\SetWidth{1.2}
\ArrowLine(12.5,0)(50,0)
\ArrowLine(50,0)(92.5,0)
\ArrowLine(92.5,0)(125,0)
\ArrowLine(125,0)(162.5,0)
\Gluon(50,3)(50,25){4}{3}
\Gluon(125,3)(125,25){4}{3}
\CCirc(50,0){3}{Black}{Yellow}
\CCirc(125,0){3}{Black}{Yellow}
\CCirc(50,25){5}{Black}{Red}
\CCirc(125,25){5}{Black}{Red}
\CCirc(87.5,0){7}{Black}{Black}
\Text(25,+10)[l]{p}
\Text(70,+10)[l]{p}
\Text(100,+10)[l]{p}
\Text(140,+10)[l]{p}
\Text(87.5,+20)[l]{$\Gamma$}
\end{picture}}


\section{Introduction}

Computing matrix elements in lattice Quantum ChromoDynamics (LQCD) needs often
the computation of renormalisation constants. Indeed, even if the lattice
computation contains only $O(a^2)$ lattice artefacts, the bare quantities differ
from the continuum ones by $O(g^2) \simeq O(1/\log(a^2))$ which is
unacceptable.  Renormalisation restores the $O(a^2)$ accuracy.  It is also known
since long that these renormalisation constants need  to be computed
non-perturbatively, using LQCD techniques. 
 
 Several non-perturbative methods have been proposed. Let us here concentrate on
those based on the MOM scheme. They start from the computation of Green
functions  of quarks, gluons and  ghosts at large enough momenta in a fixed
gauge, usually the Landau gauge. This gives the renormalisation constant
$Z(\mu)$ at many values of the scale $\mu$. Assuming that our goal is to deliver
the  renormalisation constant in the $\MSB$ scheme at say 2 GeV (a typical
phenomenological scale), one  must then use results from perturbative QCD to
convert MOM into $\MSB$ and run to 2 GeV. The running of $Z_{\mathrm MOM}(\mu)$
is a very powerful testing tool:  indeed perturbative QCD is only useful if we
are in the perturbative regime, i.e. at large enough momenta. The only way to
check whether this is the case is to compare lattice data with the perturbative
running. In this framework, it turns out that this is not always the case.

Deviations from perturbative running can be analysed via Wilson operator
expansion and the Shifman-Vainshtein-Zakharov (SVZ) sum rules. 
It turns out that the dominant non-perturbative  correction in
Landau  gauge is due to the non vanishing vacuum expectation value of the
only dimension-two operator~:  
$A^2 \equiv A^a_\mu A^{a\mu}$~\cite{Lavelle:1992yh}, and that it is not 
small~\cite{Boucaud:2001st,Boucaud:2000nd,Boucaud:2005rm,Boucaud:2005xn,
Boucaud:2008gn,Blossier:2010ky,Megias:2007pq}. 
It is thus necessary to carefully  look for the possibility of such a
contribution, which appears in the OPE, as a $1/p^2$ contribution up to
logarithmic corrections. The coefficient of this  $1/p^2$ contribution is
equal to the vacuum expectation value $g^2\VEV{A^2}$ times a Wilson
coefficient that has to be computed in perturbation  theory, and has been up
to three loops for propagators~\cite{Chetyrkin:2009kh}. To argue that a
measured $1/p^2$ contribution is a continuum power correction and not a
lattice artefact, we must check that  the $1/p^2$ term in the fit scales with
lattice spacing when expressed {\it in physical units}. To further argue that
this is indeed due to $\VEV{A^2}$ we must compare the resulting $\VEV{A^2}$
from different quantities and thus check the universality of the condensates 
which is on the ground of the SVZ technology. The theory of Wilson
operator expansion is then constraining: since there exists only one dimension-two
operator, provided that it is renormalized with the same  prescription for
all these quantities, all the different estimates of $\VEV{A^2}$  should
coincide within errors, up to $1/p^4$ corrections, for a given value of $N_f$
and of the dynamical masses. Of course, its extraction needs the coefficients of the
Wilson expansion which are computable in perturbation.
To test this universality of the extracted $\VEV{A^2}$ is one of the goals of 
our program of analysing many different quark and gluon quantities obtained from lattice gauge 
configurations produced by the European twisted mass collaboration (ETMC). 
We have also applied a criterium proposed in~\cite{Martinelli:1996pk} to validate
the way we use operator expansion.
This paper makes one of the first steps in such a program.

It is worthwile also to mention that several authors 
have elaborated further on the relation between this gauge-dependent gluon 
condensate, obtained in the Landau gauge, and possible $1/p^2$-terms in gauge 
invariant quantities, and thus on the phenomelogical implications, 
mainly in connection with confinement scenarios, of such 
a dimension-two condensate~\cite{Gubarev:2000nz}. 

All this can only be done once the lattice artefacts are eliminated or at
least under control. The $O(a^2)$ artefacts can be quite large since we
consider large momenta, while finite volume artefacts are minor. There are
two main types of $O(a^2)$ artefacts: $O(a^2p^2)$ artefacts which respect the
continuum $O(4)$ rotation symmetry, and hypercubic artefacts which respect
the $H_4$ hypercubic symmetry group but not $O(4)$. The latter are effects of
the hypercubic symmetry of the lattice action. We will identify the
$O(a^2p^2)$  artefacts non-perturbatively by doing a fit of the running
$Z(\mu)$ which will include the perturbative running, the  $\VEV{A^2}$ power
correction and a term proportional to $a^2p^2$. Notice that, while the
perturbative and $\VEV{A^2}$ running contributions must approximately scale
in  physical units, the artefacts must scale in lattice units. This is an 
additional check we shall perform. 

Concerning the elimination of hypercubic artefacts, which is better done before 
the above mentioned running fit, several methods have been proposed in
literature: the democratic selection, the perturbative correction and the
non-perturbative ``egalitarian'' one (``egalitarian" because all the points are used on the same footing in this approach). 
We will discuss this in some detail later
and perform extensive comparisons. In particular we will use a new quality 
test which consists in watching to what extent the ``half-fishbone''
structure, which raw lattice results for $Z_q$ always exhibit and which
is a dramatic illustration of hypercubic artefacts, are corrected by every 
method. 

Although all the issues raised here concern all the renormalisation constants
as well as the QCD coupling constant, we will concentrate in the following
on   $Z_q$, that renormalises the quark field 
\bea
q_R = Z_q^{1/2} q_B
\eea
where $q_B$ ($q_R$) is the bare (renormalised) quark field. 
In the RI'-MOM scheme $Z_q$ is defined by
\bea\label{Zqdef}
Z_q(\mu^2=p^2) = \frac {-i} {12\, p^2} {\mathrm Tr}[S_{bare}^{-1}(p) \;\pslash]
\eea
where $S_{\mathrm{bare}}(p)$ is the bare quark propagator. 
Our goal is to compute that constant from LQCD with twisted Wilson quarks.

In~\cite{Boucaud:2003dx,Boucaud:2005rm} a study~\footnote{In~\cite{Boucaud:2005rm}
$Z_q$ was denoted $Z_\psi$} of $Z_q$ was performed from LQCD in the case
$N_f=0$ using both the overlap and Wilson clover fermions. 
In~\cite{Boucaud:2003dx} the exceptional size of hypercubic artefatcs was
stressed and a non-perturbative elimination of hypercubic artefacts
performed along the same principle as what is used here. 
In~\cite{Boucaud:2005rm} the Wilson coefficient of the
$\VEV{A^2}$ was computed up to the leading logarithm approximation and applied to 
estimate the condensate from the LQCD data. The outcome was that a
significant non-perturbative contribution from  $\VEV{A^2}$ was needed to
account for the results. Notice that we do not expect the $\VEV{A^2}$ to be similar or even
close in the cases of $N_f=0$ and $N_f=2$. 

Summarising the above discussion, we do here concentrate on $Z_q$ because we
consider it as a kind of benchmark for the following reasons:
  
\begin{itemize}
\item It has specially large hypercubic artefacts and is thus a good 
 test bed for  a correct treatment of these.
\item It has a vanishing anomalous dimension at leading order in the Landau 
gauge: it's perturbative running is thus soft. 
\item The Wilson coefficient of the $\VEV{A^2}$ condensate is rather 
large~\cite{Boucaud:2005rm,Chetyrkin:2009kh}, which is an incentive to 
look carefully for non-perturbative contributions.
\end{itemize}

In this paper, in order to test deeply the reliability of our results, 
we will compare many fits:
perturbative/non-perturbative hypercubic correction, one-window/sliding-windows
non perturbative hypercubic correction, effect of the total fitting range,
fitting separately every $\beta$ and global fit. As a consequence we will 
proceed as follows:

\begin{itemize}
\item We will recall some general formulae concerning the perturbative
and non-perturbative running in the continuum
\item describe our lattice setting
and our non-perturbative ``egalitarian'' method to eliminate  hypercubic
artefacts; 
\item  present the results concerning the  {\it perturbative} correction 
method for hypercubic artefacts and show the quality checks;
\item present the results concerning the  {\it non-perturbative} method
to correct for hypercubic artefacts  and show the quality checks, propose 
two types of fits, the sliding windows fit (SWF) and the one-window fit (OWF). 
\item We will perform the running fit on the output of  all the previously mentioned
 hypercubic corrected data, compare the results for the $g^2 \VEV{A^2}$
 for all these fits, check the scaling of $g^2 \VEV{A^2}$;
 \item  check the scaling of the $\propto a^2p^2$ artefacts;
 \item  check the lattice spacing dependence of $Z_q^{\mathrm {pert}}$, the 
  perturbative contribution to $Z_q$, $\propto g^2$; 
 \item  study the range of variation of our results
 for  $g^2 \VEV{A^2}$ from the egalitarian method with one/sliding window(s), 
 with varying fitting ranges, and the perturbative method with two 
 realisations. We extract from there the systematic uncertainty.
 \item We will join in one plot the three $\beta$'s and perform the fit of the
 running, 
 \item compare the resulting $g^2 \VEV{A^2}$ with the 
one extracted from the strong coupling constant and with quenched 
estimates, and tested our procedure according to Martinelli-Sachrajda's
criterium~\cite{Martinelli:1996pk}. 
 \item conclude.
\end{itemize}

\section{Running of $Z_q$}
\subsection{Perturbative running}
In Landau gauge $Z_q$ has a vanishing anomalous dimension to leading order, 
i.e. its running starts at $O(\alpha^2)$. 
The perturbative running has been computed up to four 
loops~\cite{Chetyrkin:1999pq} and references therein. The needed formulae 
are accessible on the web site also indicated in ref.~\cite{Chetyrkin:1999pq}.

\subsection{Wilson expansion and non-perturbative running}

 To handle non-perturbative corrections we use the theory of 
Operator Product Expansion~\cite{Wilson69} and its application in estimating 
power suppressed non-perturbative corrections via vacuum expectation 
values~\cite{SVZ}. 
In Landau gauge there exists only one dimension-2 operator 
allowed to have a vacuum expectation value: $A^2 \equiv 
\sum_{\mu=1,4}^{a=1,8} A_\mu^a A^{a\mu}$.
The Wilson coefficient of this operator has been computed  
to leading logarithm in~\cite{Boucaud:2005rm}
and extensively for all propagators
up to $O(\alpha^4)$ in~\cite{Chetyrkin:2009kh}.

\subsubsection{$\VEV{A^2}$ tree level Wilson coefficients for $Z_q$}
In order to give a hint let us just sketch the tree level calculation
of that Wilson coefficient.
\begin{figure}[h]
  \begin{center}
    \includegraphics[width=50mm]{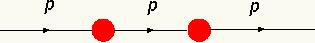}
  \end{center}
\end{figure}

Consider the above diagram, describing a quark propagating in a constant 
background gauge field. As a consequence the red bubble represents the
interaction of the quark with this background field: 
$i g {\lambda_a}/2\,  \Aslash^a$. 
The Feynman rules are then applied  as usual. Neglecting the quark mass
it gives 
$$
\frac{-i\pslash}{p^2}\left(\sum_{\mu=1, a=1}^{\mu=4, a=8}\sum_{\mu'=1, a'=1}^{\mu'=4, a'=8}i 
g \frac{\lambda_a}2 A^{a}_{\mu}\gamma^{\mu} \frac{-i\pslash}{p^2}
 i g \frac{\lambda_{a'}}2  A^{a'}_{\mu'} \gamma^{\mu'}\delta_{aa'}\delta_{\mu\mu'}\right)\frac{-i\pslash}{p^2}$$
\beq
= -\frac {g^2} {12} \frac {\VEV{A^2}}{p^2} \times \frac{-i\pslash}{p^2}
\eeq
where $\langle \, \, \rangle$ represents the vacuum expectation value, 
 $\sum \lambda_a^2/4 = C_F =4/3$ (proportional to the identity matrix in color
 space), the sum over $\mu$ gives a
factor 4, and
\bea\label{form}\langle(A_\mu^a)^2\rangle=\langle A^2/32\rangle, \qquad
\langle (A\cdot \hat p)^2 \rangle =
\langle A^2/4\rangle\eea
from the homogeneity of the vacuum for rotations in space-time and
color space.

For $Z_q$ defined by~\eq{Zqdef}, we get at tree level the following
non-perturbative  contribution due to $\VEV{A^2}$:
\bea
\delta Z_q =  \frac {g^2} {12} \frac {\VEV{A^2}}{p^2}
\eea

\subsubsection{The Wilson coefficients at $O(\alpha^4)$}
The Wilson coefficient of $\VEV{A^2}$ for the quark propagator has been computed
up to $O(\alpha^4)$ in~\cite{Chetyrkin:2009kh} in the $\overline{\rm MS}$ scheme. 
Our lattice data refer naturally to the RI'-MOM scheme. Some work is needed to
derive the correct analytic formula which allows a fitting of our lattice data.
We have derived this in the appendix~\ref{appendix2}.


\section{The lattice computations}
\label{sec:lat}

The results presented here are based on the gauge field 
configurations generated by the European Twisted Mass Collaboration 
(ETMC) with the tree-level improved Symanzik gauge action~\cite{Weisz:1982zw}  
and the twisted mass fermionic action~\cite{Frezzotti:2000nk} at
maximal twist. 

\subsection{The lattice action}

A very detailed discussion about the twisted mass and tree-level improved
Symanzik  gauge actions, and about the way they are implemented by ETMC, can be
found in 
refs.~\cite{Boucaud:2007uk,Boucaud:2008xu,Urbach:2007rt,Dimopoulos:2008sy}. 
Here, for the sake of completeness, we will present a brief reminder of the
twisted action and the run parameters for the gauge configurations that will be
exploited in the present work (See tab.~\ref{setup}).

The Wilson twisted mass fermionic lattice action for two flavours of mass
degenerate  quarks, reads  (in the so called twisted
basis~\cite{Frezzotti:2000nk,Frezzotti:2003ni} ) 
\begin{align}
  \label{eq:Sf} 
  \begin{split} 
    S_\mathrm{tm}^{\rm F} = &\, a^4\sum_x\Bigl\{ 
    \bar\chi_x\left[D_{\rm W}+ m_0 + i\gamma_5\tau_3\mu_q  
    \right]\chi_x\Bigr\}\, , \\ 
    & D_{\rm W} = \frac{1}{2}\gamma_\mu\left(\nabla_\mu+\nabla_\mu^{*}\right) 
    -\frac{ar}{2}\nabla_\mu\nabla_\mu^{*} \, ,
  \end{split} 
\end{align}
where $m_0$ is the bare untwisted quark mass and $\mu_q$ the bare twisted  quark
mass, $\tau_3$ is the third Pauli matrix acting in flavour space   and $r$ is
the Wilson parameter, which is set to $r=1$ in the simulations. The twisted
 Dirac operator is defined as
\bea\label{eq:DIRtw}
D_{\rm tw} \equiv D_{\rm W} + m_0 + i\gamma_5\tau_3\mu_q \ .
\eea
The operators
$\nabla_\mu$ and $\nabla_\mu^{*}$ stand for the gauge covariant nearest  
neighbour forward and backward lattice derivatives:  
\bea\label{def}
\nabla_\mu(x,y) &\equiv&\left[ \delta_{y,x+\hat \mu}\,U_\mu(x) - \delta_{x,y}\right] \ , \no
\nabla^\star_\mu(x,y)&=&\left[\delta_{x,y} -
\delta_{y,x-\hat \mu}\,U^\dagger_\mu(x-\hat \mu)\right] \ , \no
D_\mu \equiv \frac 12\left [\nabla_\mu(x,y)+\nabla^\star_\mu(x,y)\right]&=&\frac 12 \left[\delta_{y,x+\hat \mu}\,W(x,y) - 
\delta_{y,x-\hat \mu}\,W(x,y)\right] \ ;
\eea 
defining the operator $D_\mu$ as the discretized covariant derivative.
The bare quark mass $m_0$
is related as usual to the so-called hopping   parameter $\kappa$, by
$\kappa=1/(8+2am_0)$. Twisted mass fermions are said to be at   {\em maximal
twist} if the bare untwisted mass is tuned to its critical   value,
$m_\mathrm{crit}$. This is in practice done by setting the so-called untwisted
PCAC mass to zero.

In the gauge sector  the  tree-level Symanzik improved 
gauge action (tlSym)~\cite{Weisz:1982zw} is applied. This action includes besides the 
plaquette term $U^{1\times1}_{x,\mu,\nu}$ also rectangular $(1\times2)$ Wilson loops 
$U^{1\times2}_{x,\mu,\nu}$. It reads  
\beq 
  \label{eq:Sg} 
    S_g =  \frac{\beta}{3}\sum_x\Biggl(  b_0\sum_{\substack{ 
      \mu,\nu=1\\1\leq\mu<\nu}}^4\{1-\re\tr(U^{1\times1}_{x,\mu,\nu})\}\Bigr.  
     \Bigl.+ 
    b_1\sum_{\substack{\mu,\nu=1\\\mu\neq\nu}}^4\{1 
    -\re\tr(U^{1\times2}_{x,\mu,\nu})\}\Biggr)\, , 
\eeq
where $\beta \equiv 6 / g_0^2$, $g_0$ being the bare lattice  coupling and it
is set    $b_1=-1/12$ (with $b_0=1-8b_1$ as dictated by the requirement  of
continuum limit normalization). Note that at $b_1=0$ this action becomes the
usual Wilson plaquette gauge action.  The run parameters for $\beta$ and $\mu_q$
of the gauge configurations that will  be exploited in the following can be
found in tab.~\ref{setup}.
\begin{table}[ht]
\centering
\begin{tabular}{||c|c|c|c|c|c|c||}
\hline
\hline
$\beta$ & $a$ fm &  $a^{-1}$ GeV &$a \mu_q$ & Volume & \# confs 
\\ \hline
$3.9$ &  0.083&2.373&
0.004 
& 
$24^3\times48$ & 
$100$ 
\\ \hline
$4.05$ &  0.0675&2.897&
0.006 &
 $24^3\times48$  
&
$100$ 
\\ \hline
$4.2$ & 0.055&3.58
 & 0.002  &
 $24^3\times48$ 
&
$100$
\\ \hline
\hline
\end{tabular}
\caption{Run parameters of the exploited data from ETMC collaboration for the
present study of $Z_q$. The second column lists the lattice spacings which we
have used in this study. It is easy to  convert it to other lattice spacings.}
\label{setup}
\end{table}

\subsection{The computation of the quark propagator}

Computing the renormalisation constants for the quark propagator and the
operators containing quark fields  demands to compute first the gauge-fixed
2-point quark  Green functions from the lattice. We exploited ETMC
gauge configurations~\cite{Baron:2009wt} obtained for  $\beta=3.9$, $\beta=4.05$ and $\beta=4.2$.
After checking the small dependence of $Z_q$ on the dynamical and valence
quark  masses we decided to use only one mass for every $\beta$,
table~\ref{setup}. The lattice gauge configurations are transformed to Landau
gauge by minimising the following functional of the SU(3) matrices, $U_\mu(x)$,
\beq
F_U[g] = \mbox{\rm Re}\left[ \sum_x \sum_\mu  \hbox{Tr}\left(1-\frac{1}{N}g(x)U_\mu(x)g^\dagger(x+\mu) \right) \right] \ ,
\eeq
with respect to the gauge transform $g$, by applying a combination of
overrelaxation algorithm and  Fourier acceleration~\footnote{We end when 
$|\partial_\mu A_\mu|^2 <10^{-11}$ and when the spatial integral of $A_0$ is
constant in time to better than $10^{-6}$.}. 

We compute quark propagators with a local source taken at a random point $x_0$
on the lattice, in order to reduce the correlation between successive
configurations:
\bea
S(y,x_0)^{a,\alpha;b_0,\beta_0}_{j} = D_{\rm tw}^{-1}(y,x)^{a,\alpha;b,\beta;i,j} 
so^{b,\beta}_{j}(x,x_0) \ , \ \qquad so^{b,\beta}_{j}(x,x_0) = \delta_{x,x_0} \delta_{b,b_0}
\delta_{\beta,\beta_0} \ ; 
\eea
where  the equation is solved for every 
$b_0=1,3$ and $\beta_0=1,4$, and $j=u,d$ labels the isospin.
 We perform the Fourier transform which is a $12\times12$ complex matrix
 \bea\label{Sp}
 S_i(p) \equiv \sum_{y} e^{-i p (y-x_0)} \,S_i(y,x_0) \ .
 \eea
This is the Fourier transform of the quark incoming to the source (the arrow
pointing towards the source). The Fourier transforms of the quark outgoing from 
the source is 
\bea
S^{\dagger 5}_{i}(p) = \gamma_5 S^{\dagger}_{\bar i}(p) \gamma_5 \ ,
\eea
where $\bar u\equiv d; \bar d\equiv u$.
From~\eq{Zqdef} the lattice quark renormalisation constant $Z_q$ is 
given by
\bea\label{Zq}
Z_q(p)\equiv \frac {-i} {12\,\tilde p^2} \;<{ \rm Tr} [S^{-1}(p) \,\tilde \pslash]> \ , 
\eea
where $<...>$ means here the average over the chosen ensemble of thermalised
configurations and $\tilde{p}_\mu = \frac{1}{a} \sin ap_\mu$. The reason to use 
$\tilde{p}_\mu = \frac{1}{a} \sin ap_\mu$ is to get $Z_q=1$ for a free fermion,
or in other words, to eliminate hypercubic artefacts at tree level. 

\subsection{The method of non-perturbative Hypercubic $H(4)$ correction}
\label{NPhyp}

The lattice estimates of the quark field renormalisation constant 
and the vertex functions lead to dimensionless quantities that, because of 
general dimensional arguments, depend on the strong interaction scale 
$\Lambda_{\rm QCD}$ and on the lattice  momentum $a\,{p}_\mu$.
%
%
%
We have computed the Fourier transforms for the following momenta:
\beq\label{pmu}
p_i = \frac{2\pi n_i}{N_L a} \qquad n_i=-N_L/4,\cdots,N_L/4 \ , \ \qquad p_4= \frac{\pi (2
n_4 + 1)}{N_T a}\qquad n_4=-N_T/4,\cdots,N_T/4 \ ; \nonumber \\
\eeq
 where $p_i=1,3$ are the spatial momenta and  $p_4$ the time like. The
 antiperiodic boundary condition in the time direction explains the $\pi (2
n_4 + 1)$ factor. 

The lattice action \eq{eq:Sf} and \eq{eq:Sg} is invariant under the 
hypercubic group $H(4)$. However the boundary conditions and the difference 
between the spatial size $N_L$ and the time-like one $N_T=2 \,N_L$ generate
finite volume corrections to the hypercubic symmetry. Only the cubic symmetry
is exact. We define cubic invariant quantities and compute their average over
the cubic group. We have thus a set of measures for every orbit of the cubic
group, labelled by 
\bea\label{cubic} 
\left( \sum_{i=1,3} p_i^m, \;\; p_4 \right) \ ,
\eea 
where $m=2,4,6$.

A first kind of artefacts that can be systematically 
cured~\cite{Becirevic:1999uc,Boucaud:2003dx,deSoto:2007ht} are those due to the breaking of
the  rotational symmetry of the Euclidean space-time when using an hypercubic
lattice,  where this symmetry is restricted to the discrete hypercubic $H(4)$
isometry group. However, as already mentioned, we have also finite volume
effects which break $H(4)$. We therefore need to adapt the 
method. One idea could be to generalise it to a cubic symmetry. This happens not
to be practical due to too few cubic symmetric orbits for a given $\vec \! p^2$.
We choose another approach motivated by the fact that the lattice action is
indeed  $H(4)$ symmetric and that finite volume effects are expected to be 
small at large momenta
compared to  finite lattice spacing artefacts. We therefore use a slight
variation of the method
described in~\cite{Becirevic:1999uc,deSoto:2007ht} {\it : we apply it to the
cubic orbits of~\eq{cubic}}, keeping track of $p_4$ which is not an $H(4)$
symmetric quantity.  

Defining the $H(4)$ invariants
 \beq
 p^{[4]}=\sum_{\mu=1}^{4} p_\mu^4 \ , \qquad p^{[6]}=\sum_{\mu=1}^{4} p_\mu^6 \ ,
 \qquad p^{[8]}=\sum_{\mu=1}^{4} p_\mu^8 \ ;
 \eeq
it happens that every cubic orbit~\eq{cubic} has a well defined set of
values for these $H(4)$ invariants, but several cubic orbits may have the 
same $H(4)$ invariants. We will neglect $p^{[8]}$ which plays no role on small
lattices.
 We can thus define the quantity $Z_q(a p_\mu)$ averaged over 
the cubic orbits as
\beq\label{Q246}
Z_q^{\mathrm {latt}}(a^2\,p^2, a^4p^{[4]}, a^6 p^{[6]}, 
ap_4, a^2\Lambda_{\rm QCD}^2) \ .
\eeq

We expect the hypercubic effects to be $O(a^2)$ lattice artefacts and therefore
to be expandable into powers of $a^2$. This would of course trivially be the
case if $a^2\,p^2 \ll 1$ since then, for example  $\epsilon=a^2 p^{[4]}/p^2 \le
a^2\,p^2 \ll 1$ (we take on purpose  this quantity which will be seen to be
dominant). Then a Taylor expansion of  \eq{Q246} will ensure the artefact to be 
$O(a^2)$. However, aiming at measuring $Z_q$ at large momentum we go up to
$a^2\,p^2 \sim 3 - 4$. We will assume, and then check, that the $Z_q^{\mathrm
{latt}}$ in \eq{Q246} can be Taylor-expanded around $p^{[4]}=0$  up to
$\epsilon$ significantly larger than 1: %
\beq\label{eq:p4expan}
Z_q^{\mathrm {latt}}(a^2\,{p}^2, a^4p^{[4]}, a^6 p^{[6]}, ap_4, a^2\Lambda_{\rm
QCD}^2) &=& Z_q^{\mathrm {hyp\_corrected}}(a^2p^2, ap_4, a^2\Lambda_{\rm QCD}^2) 
\nonumber \\ 
&+&
R(a^2p^2,a^2\Lambda_{\rm QCD}^2) \, a^2 \frac{p^{[4]}}{p^2} \ + \
\cdots 
\eeq 
where
\beq\label{eq:deriv}
\left.R(a^2p^2,a^2\Lambda_{\rm QCD}^2) =  \frac{dZ_q^{\mathrm
{latt}}\left(a^2p^2 ,0,0,0,a^2\Lambda_{\rm
QCD}^2\right)}{d\epsilon}\right|_{\epsilon=0}.
\eeq
Of course terms proportional to $p^{[6]}$, $p^{[4]2}$, etc. can be added
analogously to the formula, as well as terms breaking 
$H(4)$. However we have found that our data were not accurate enough
to allow fitting them, and that using only \eq{eq:p4expan} and \eq{eq:deriv} gave
satisfactory fits. 
Now we must describe how we fit the functions appearing in the r.h.s of 
\eq{eq:p4expan}.

\subsubsection{The sliding window fit (SWF)} 
\label{sec:slid}
We consider all values of $a^2p^2$ in a range:  $a^2p_{\mathrm{min\_in}}^2 \le
a^2p^2 \le a^2p_{\mathrm{max\_in}}^2$, each of which contains a set of cubic
orbits.  We choose an integer width $w$ (we will use $w=10$ in numerical
applications) and define a
window as  the set of $2w + 1$ values of $a^2p^2$ around a 
$a^2p^2_{\mathrm {center}}$ (
$w$ contiguous values  below  $a^2p^2_{\mathrm {center}}$ and as many above). 
There are as many
windows as values of  $a^2p^2_{\mathrm {center}}$ such that all of them are
in the range  $[a^2p_{\mathrm{min\_in}}^2,a^2p_{\mathrm{max\_in}}^2]$. 
This defines the
range of interest $a^2p_{\mathrm{min\_out}}^2 \le  a^2p^2_{\mathrm {center}} \le
a^2p_{\mathrm{max\_out}}^2$.

For every window we use for the fit all cubic orbits corresponding to the values
of $a^2p^2$ in the window. We fit, according to \eq{eq:p4expan}, $2\, w+2$
parameters which are the $2\, w+1$ values of  $ Z_q^{\mathrm
{hyp\_corrected}}(a^2p^2,a^2\Lambda_{\rm QCD}^2)$ within the window, and one
common value of $R(a^2p^2_{\mathrm center},a^2\Lambda_{\rm QCD}^2)$. The
dependence in these parameters is linear and thus the fit amounts to invert a
matrix. It is clear that for any $a^2p^2$ the  $ Z_q^{\mathrm
{hyp\_corrected}}(a^2p^2,a^2\Lambda_{\rm QCD}^2)$ is fitted every time $a^2p^2$
is within a window, i.e. $2\, w+1$ times. We keep as the final result only the
result of the fit when $a^2p^2$ is the center of the window. At the end of the
fit, for every $a^2p^2_{\mathrm {center}}$  in the range 
$a^2p_{\mathrm{min\_out}}^2 \le a^2p^2_{\mathrm center} \le
a^2p_{\mathrm{max\_out}}^2$ we have, as expected, a fitted value for both
functions of the r.h.s of \eq{eq:p4expan}. 

We can then study the function $R(a^2p^2,a^2\Lambda_{\rm QCD}^2)$.
As will be reported later (see for instance \fig{fig:slopes}) we find that
a reasonable approximation for $R$ is
\bea\label{eq:ow}
R(a^2p^2,a^2\Lambda_{\rm QCD}^2)= c_{a2p4} + c_{a4p4} \, a^2p^2 \ .
\eea
This leads to the one window fit.

\subsubsection{The one window fit (OWF)}
\label{sec:one}
We tune $w$ such that only one (or at worst two) windows are included in the
range $[a^2p_{\mathrm{min\_in}}^2,a^2p_{\mathrm{max\_in}}^2]$. We then perform
the fit for that window according to the equation
\bea\label{eq:owfexpan}
Z_q^{\mathrm {latt}}(a^2\,{p}^2, a^4p^{[4]}, a^6 p^{[6]}, ap_4, a^2\Lambda_{\rm
QCD}^2) &=& Z_q^{\mathrm {hyp\_corrected}}(a^2p^2,a^2\Lambda_{\rm QCD}^2) +
c_{a2p4} \, a^2 \frac{p^{[4]}}{p^2} \no
&+& c_{a4p4} \ a^4 p^{[4]} \ .
\eea
This fit gives $2 w +3$ parameters which are 
$Z_q^{\mathrm {hyp\_corrected}}(a^2p^2,a^2\Lambda_{\rm QCD}^2)$ for all 
$a^2p^2$ in the window, i.e. in the range
 $[a^2p_{\mathrm{min\_in}}^2,a^2p_{\mathrm{max\_in}}^2]$, (or if the range size
 is even one value is eliminated) and the parameters $c_{a2p4}$ and $c_{a4p4}$.

\subsection{Other lattice artefacts} 

There are ultraviolet artefacts which are  functions of $a^2\,p^2$ and are thus
insensitive to hypercubic biases and not corrected by the above-mentioned
method. They will be corrected simply by assuming a term linear in $a^2p^2$ in
the final fit and check that the coefficient scales correctly for different
lattice spacings.

To take into account the space-time anisotropy, which is a finite volume
artefact, we can check the dependence of $Z_q$  on a anisotropic quantity such
as $p_4^2-\vec\, \! p ^2/3$.  We did not see any sizeable effect of this
parameter.

Finite volume artefacts are also studied as usual by a comparison of runs at
different volume. We expect a small effect at large momenta and our checks
confirmed it, as well as the analysis of~\cite{Blossier:2010ky}. We will not
consider this artefact anymore.

\section{Lattice results and hypercubic corrections} 
\label{sec:Zqhyp}

\begin{figure}[h]
  \begin{center}
    \includegraphics[width=10cm]{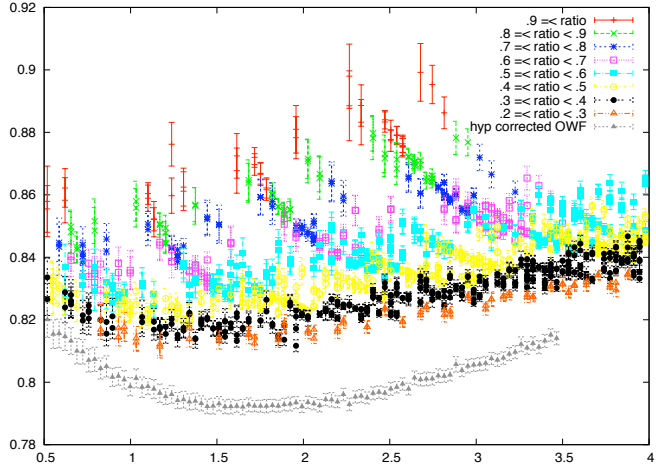}
  \end{center}
\caption{\small This plot shows the raw data for $\beta=3.9$, $Z_q^{\mathrm
{latt}}(a^2\,p^2, a^4p^{[4]}, a^6 p^{[6]},  ap_4, a^2\Lambda_{\rm QCD}^2)$ in
\eq{eq:p4expan}, in terms of $a^2 p^2$ in the horizontal axis. 
The ``half-fishbone structure'' due to hypercubic artefacts is
clearly seen. There is one point for every cubic (3-D) orbit. The color code 
classifies the data according to their degree of ``democracy'' measured by 
$\mathrm{ratio}=p^{[4]}/(p^2)^2 $. The lowest plot corresponds to the
non-perturbatively hypercubic corrected data (or ``egalitarian result'') resulting 
form the one-window fit $Z_q^{\mathrm
{hyp\_corrected}}(a^2p^2,a^2\Lambda_{\rm QCD}^2)$ in \eq{eq:owfexpan}.
 The  data correspond to  $\beta=3.9, a\mu=0.004$,
but the same features appear for all $\beta$'s. }
\label{fig:fishbone}
\end{figure}

The hypercubic artefacts generate on the raw lattice data,  $Z_q^{\mathrm
{latt}}$, the so called ``half-fishbone''  structure~\cite{Boucaud:2003dx} shown
in fig~\ref{fig:fishbone}. In this figure all the points labelled as explained
in \eq{pmu} and \eq{cubic} are plotted.  The color code shows the value of the
ratio $p^{[4]}/(p^2)^2$ which is  between 0.25 and 1. The closer to 1 means the
less ``democratic'' or ``tyrannic'' ones. We see, as expected,  that the tyrannic  points
are more affected by the artefacts. We also see that the gap between
$Z_q^{\mathrm {latt}}(a^2\,p^2, a^4p^{[4]}, a^6 p^{[6]},  ap_4,  a^2\Lambda_{\rm
QCD}^2)$ at a given $  p^2$ can be as large as 0.07, i.e; about 10\%. Taking a
naive average, without a correct treatment  of this artefact  would leave a
systematic upward shift of about 5 \%. In section \ref{NPhyp} we have developed
a non-perturbative method to cure this artefact (this method being indeed
splited in two options, the SWF and the OWF). 
There exist two other methods.
The oldest one is the ``democratic selection''. It amounts to  keep only, say,
the points with $\mathrm{ratio} \le 0.3$ in  fig~\ref{fig:fishbone}. One sees
that it eliminates a lot of data and it is still far from the ``egalitarian''
 result (the lowest curve in fig~\ref{fig:fishbone}) which we consider as
 better.  We will thus not consider further this democratic selection. 
The second other method to correct hypercubic artefacts uses perturbative calculation
and is detailed in the next section.

\subsection{Perturbative correction} 
\label{sec:pert}

\begin{figure}[hbt]
\begin{center}
\begin{tabular}{cc}
\includegraphics[width=9.7cm]{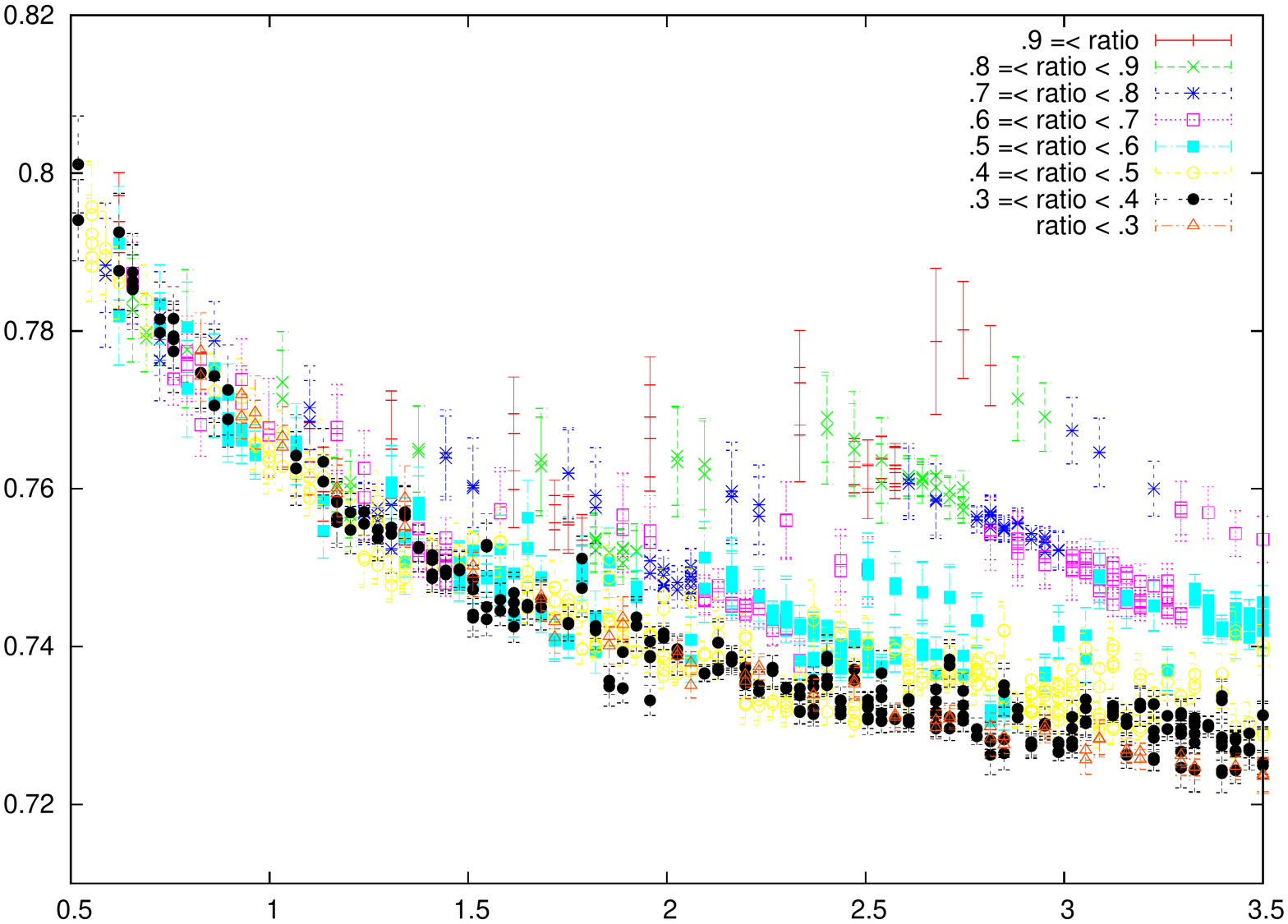}
&
\hspace*{-1.75cm}
\includegraphics[width=9.7cm]{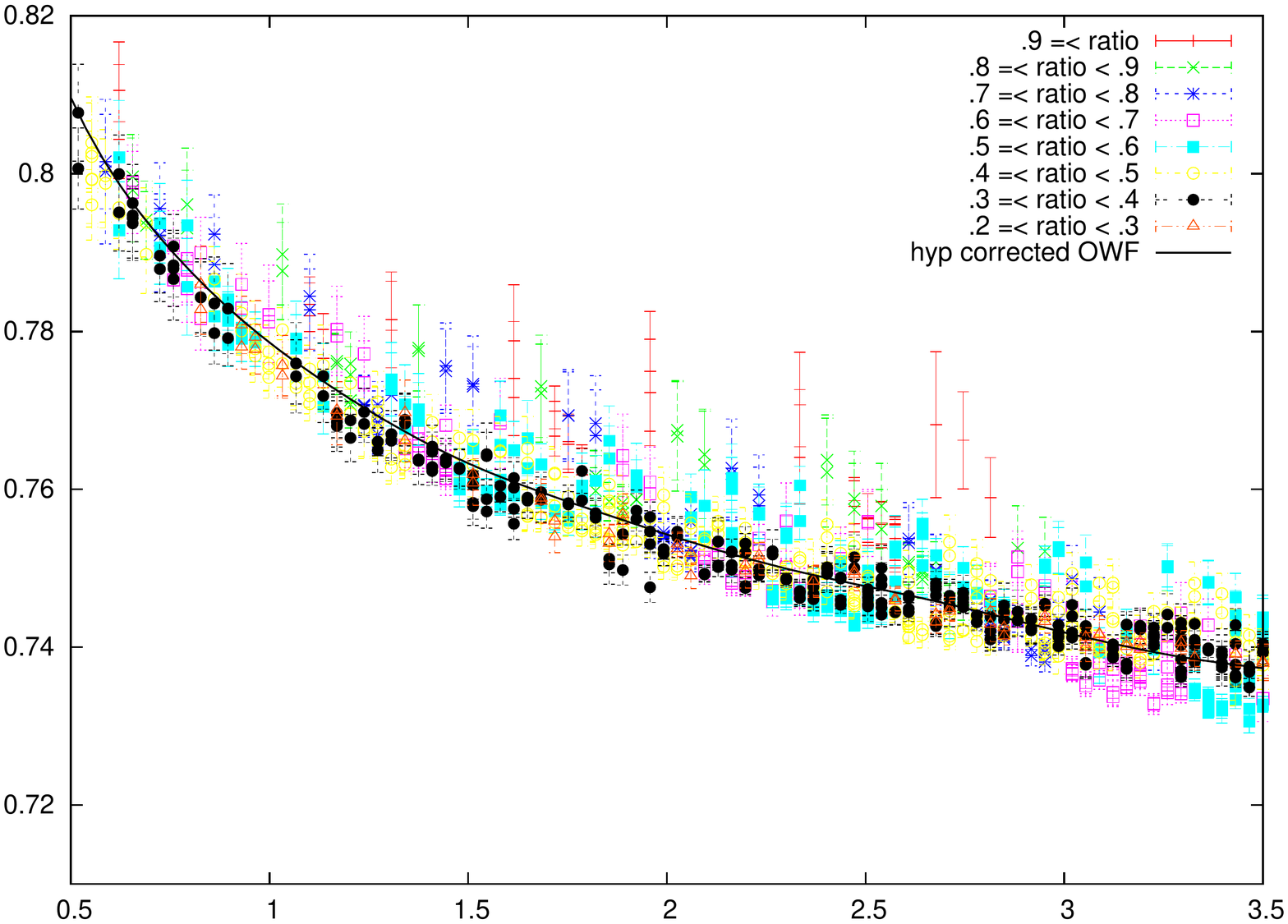}
\end{tabular}
\end{center}
\caption{\small  L.h.s: data of fig.~\ref{fig:fishbone}
corrected by the perturbative subtraction, formula of~\eq{eq:subpertild},
section~\ref{sec:subpertild}. 
 R.h.s: same exercice with the prescription \eq{eq:subpert}
 with \eq{eq:feli}. The color code is the same as
 in fig~\ref{fig:fishbone} and $\mathrm{ratio}=p^{[4]}/(p^2)^2 $.
The horizontal axis is $a^2 p^2$ for both.}
\label{fig:pert}
\end{figure}

 The perturbative method~\cite{Constantinou:2009tr} consists in computing at one loop in lattice
perturbation theory~\cite{Capitani:2002mp},  then, assuming that the lattice spacing
artefacts are reliably described by  the $O(g^2 a^2)$ terms thus obtained, in
subtracting them from the lattice data. This method has been applied to quark
bilinear operators in~\cite{Constantinou:2010gr}. For comparison  we have
applied here  this method following the prescription  described in section 3.2.2
of~\cite{Constantinou:2010gr} and also a variant of it. 

Eq (24) in~\cite{Constantinou:2009tr} may be written in Landau gauge as
\bea\label{eq:pert}
Z_q^{\mathrm pert}(a^2p^2)&=&Z_q^{\mathrm tree}(a^2p^2)
+ \tilde g^2\big(b_{q1}  + 
c_{q1} \; a^2p^2 + c_{q2}\; a^2p^2\log(a^2p^2) + \no
&&c_{q3} \frac{a^2 p^{[4]}}{p^2} + c_{q4} \frac{a^2 p^{[4]}}{p^2} \log(a^2p^2)\big) \ ,
 \eea
 where, to follow~\cite{Constantinou:2010gr}, $\tilde g ^2 = g_{\mathrm
 {boosted}}^2/(12\pi^2)$, with $g_{\mathrm
 {boosted}}^2 = g_{\mathrm {bare}}^2/\mathrm {<plaquette>}$.
The coefficients are  defined  using the notations 
of eq. (24)  in~\cite{Constantinou:2009tr}:
\bea\label{eq:feli}
\begin{array}{ll}
\displaystyle
c_{q1} = \epsilon^{(2,4)} \ , & \displaystyle c_{q2} = \frac{59}{240}+ \frac {c_1}{2} + \frac
{C_2}{60} \ , \nonumber \\
\displaystyle c_{q3} = \epsilon^{(2,1)} - \frac{3}{80} -
\frac{C_2}{10} \ , & \displaystyle \rule[0cm]{0cm}{0.9cm} c_{q4} = \frac{101}{120} -\frac{11}{30}\,C_2 \ .
\end{array}
\eea

\subsubsection{Prescription with $\tilde p_\mu$}
\label{sec:subpertild}

Using the prescription of eq.~(35) in~\cite{Constantinou:2010gr}, for every
cubic orbit we define  the substracted quantity as~:
\bea\label{eq:subpertild}
Z_q^{\mathrm {pert\_tilde}}(a^2\,{p}^2, a^4p^{[4]}, a^6 p^{[6]}, ap_4, a^2\Lambda_{\rm
QCD}^2) \ = \ Z_q^{\mathrm {latt}}(a^2\,{p}^2, a^4p^{[4]}, a^6 p^{[6]}, ap_4, a^2\Lambda_{\rm
QCD}^2) \no 
- \  \tilde g^2\left( c_{q1} \; a^2\tilde p^2 + c_{q2}\; a^2\tilde p^2\log(a^2\tilde p^2) + 
c_{q3} \frac{a^2 \tilde p^{[4]}}{\tilde p^2} + c_{q4} 
\frac{a^2\tilde  p^{[4]}}{\tilde p^2} \log(a^2\tilde p^2) \right)
\eea 
The result of this substraction is plotted in the  l.h.s of fig.~\ref{fig:pert}.
The half-fishbone structure is still clearly visible
for $a^2p^2 > 1.6$. We then try another prescription.

\subsubsection{Prescription with $p_\mu$}
\label{sec:subpert}

The trace of \eq{Zq}, which introduces $\tilde p_\mu$, had to be applied 
in eq.~(24) of~\cite{Constantinou:2009tr} to obtain \eq{eq:pert}. 
We will now expand in $p_\mu$ before performing the trace and then keep the 
 $O(g^2 a^2)$ terms for substraction. This gives, using 
 again~\cite{Constantinou:2009tr},
\bea\label{eq:subpert}
Z_q^{\mathrm {pert\_notilde}}(a^2\,{p}^2, a^4p^{[4]}, a^6 p^{[6]},
 ap_4, a^2\Lambda_{\rm
QCD}^2) \ = \ Z_q^{\mathrm {latt}}(a^2\,{p}^2, a^4p^{[4]}, a^6 p^{[6]}, ap_4, a^2\Lambda_{\rm
QCD}^2)\no 
-  \ \tilde  g^2 \left( 
c_{q1} \; a^2 p^2 + c_{q2}\; a^2 p\log(a^2 p^2) + 
c_{q3}' \frac{a^2  p^{[4]}}{ p^2} + c_{q4} 
\frac{a^2  p^{[4]}}{ p^2} \log(a^2 p^2) \right) \ ,
\eea 
 where
\bea\label{eq:feli2}
\quad c_{q3}'  \ = \ \frac {\epsilon^{(0,1)}} 6 + \epsilon^{(2,1)} - \frac{3}{80} -
\frac{C_2}{10} \ = \ c_{q3} + \frac {\epsilon^{(0,1)}} 6 \ .
\eea 
This result is plotted in the r.h.s. of fig.~\ref{fig:pert}. With this variant, the half-fishbone is
significantly reduced  but the dominant $O(4)$ artefact is overcorrected : a  linear  behaviour in
$a^2p^2$ is clearly visible, larger than with  the first prescription.

\subsubsection{Lessons about the perturbative method}
We see that the two prescriptions start
differing significantly at $a^2p^2 \simeq 1$, which is not surprising since
higher order terms become significant, for example, an  $a^4 p^{[4]}$ is also of
order 1 for tyrannic points, while  $a^2 p^2 - a^2  \tilde p^2 \simeq 0.3
$. The perturbative method goes in the right direction, but it is impossible to
know a priori its quality without performing the tests we propose here. The
method contains several ambiguities: what to take for the coupling constant ?
use $p_\mu$ or $\tilde p_\mu$? Contrarily to the non-perturbative method it 
provides both the hypercubic corrections and the $O(a^2p^2)$ ones. Conceptually 
the perturbative method is very useful as it exhibits qualitative aspects which
may guide the use of the non-perturbative one, for example it justifies the 
smoothness of the variation of the derivative $R$~\eq{eq:deriv} as a function of
$a^2p^2$ as well as that of the slope in $a^2p^2$. Finally we shall see that it
gives results similar to the non-perturbative ones. 

\subsection{Non perturbative hypercubic correction}
\subsubsection{Sliding window fit versus one window fit}
\begin{figure}[hbt]
  \begin{center}\begin{tabular}{cc}
    \includegraphics[width=97mm]{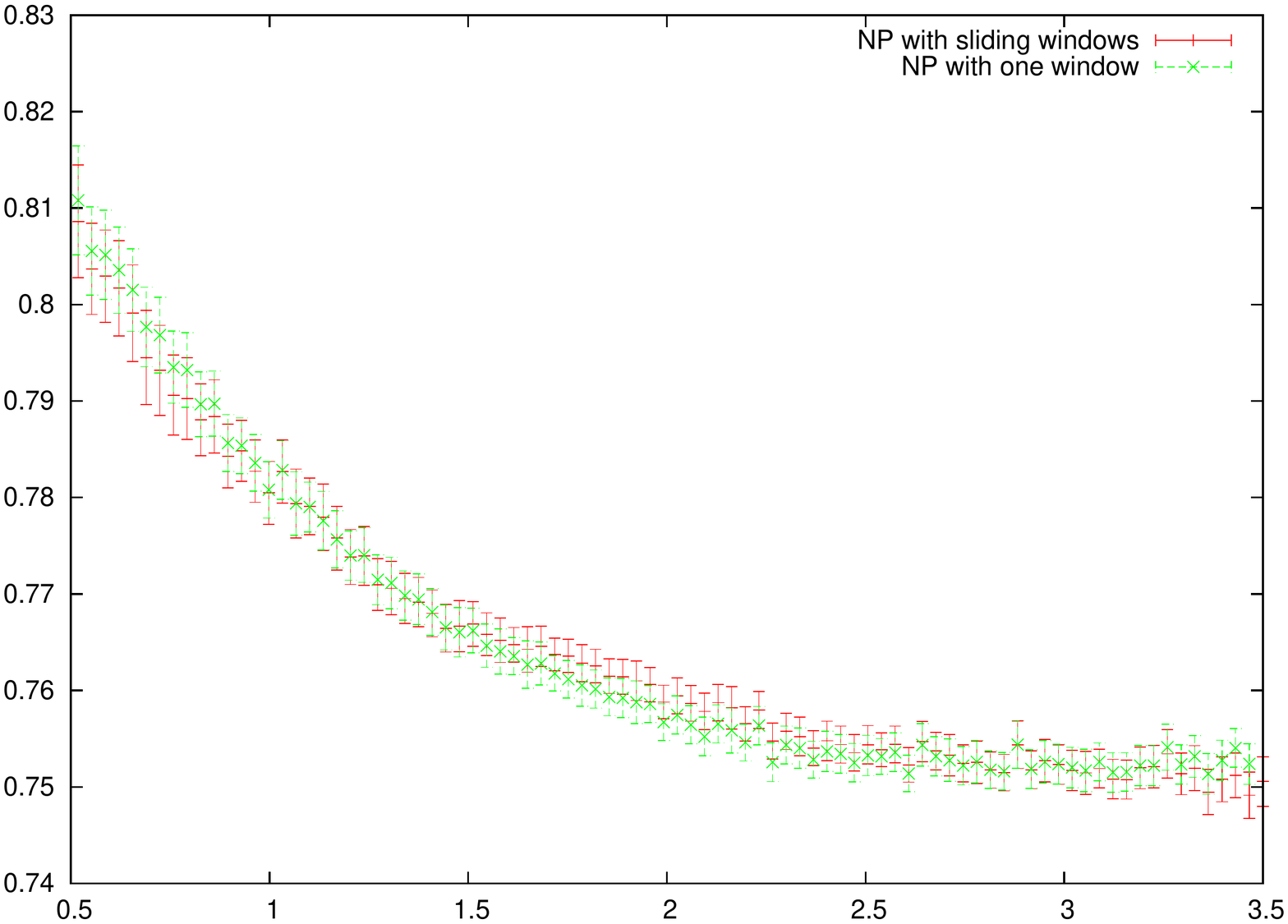}
	&
\hspace*{-1.75cm}
    \includegraphics[width=97mm]{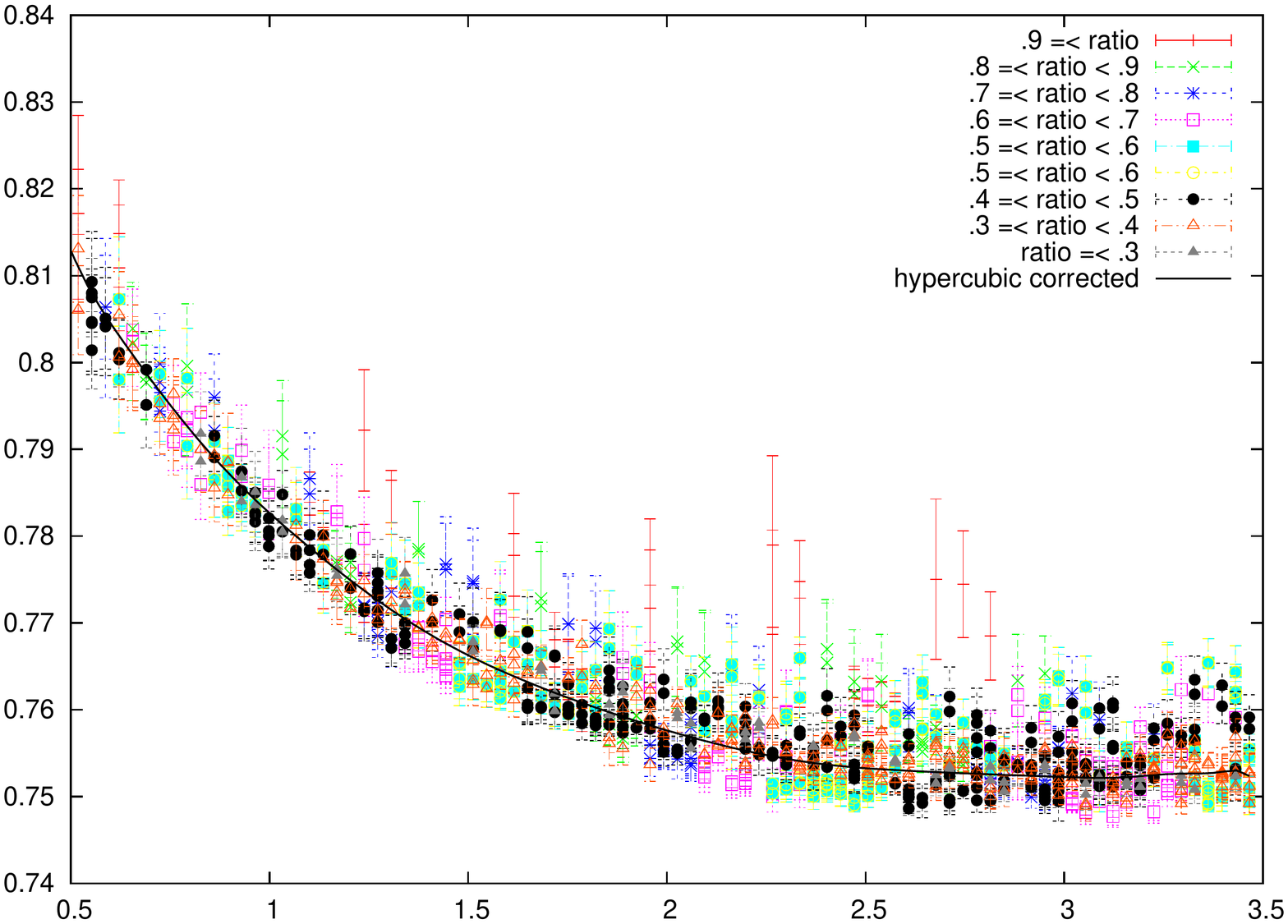}	
  \end{tabular}\end{center}
\caption{\small On the l.h.s we compare for $\beta=3.9, a\mu=0.004$
the non-perturbatively  corrected $Z_q$ from the sliding-window fit (SWF),
section~\ref{sec:slid}  and from the one-window one (OWF), section~\ref{sec:one}. 
To show better the running we have also
subtracted the  $O(a^2p^2)$ artefact which will be computed below. On
the r.h.s we show, using the OWF, the non-perturbatively subtracted
data~\eq{eq:NPsub}. 
 There is one point for every cubic (3-D) orbit. The
color code and the definition of the parameter ratio  are the same as  in fig.~\ref{fig:fishbone}.
 The black line
corresponds to the OWF  non-perturbatively corrected result:
$Z_q^{\mathrm {hyp\_corrected}}(a^2p^2,a^2\Lambda_{\rm QCD}^2)$ of 
\eq{eq:owfexpan}. The $O(a^2p^2)$ has also been subtracted.
The horizontal axis is $a^2 p^2$.}
\label{fig:NP_subtracted}
\end{figure}

We now apply the non-perturbative correction. We only
expand in $p^{[4]}$ since the higher order terms turn out to be
negligible in our momentum range. In section~\ref{NPhyp} we have presented two types of fits, similar
in spirit: the sliding window fit (SWF) described in section~\ref{sec:slid} 
which amounts to using \eq{eq:p4expan} combined with \eq{eq:deriv} and the one
 window fit (OWF) described in section~\ref{sec:one}  which amounts to using
  \eq{eq:ow}. In the l.h.s of \fig{fig:NP_subtracted}  we show, in the case 
of $\beta=3.9$, the comparison of hypercubic corrected data after applying
OWF and SWF. The difference does not appear to be large which is rather
encouraging. The OWF gives a slightly smoother result. For this value of $\beta$ the chi-squared
is not good (see table~\ref{tab:p4slopes}) but remember it uses
only two hypercubic parameters. Chi-squared for the other $\beta$'s are better.

\subsubsection{Half-fishbone reduction test}
We need also to apply the half-fishbone reduction test as in the
perturbative case, i.e. to subtract to the raw data of every cubic orbit the
hypercubic correction. We present the OWF result. From \eq{eq:owfexpan} the
subtraction amounts to 
 \bea\label{eq:NPsub}
Z_q^{\mathrm {non-pert\_OWF}}(a^2\,{p}^2, a^4p^{[4]}, a^6 p^{[6]},
 ap_4, a^2\Lambda_{\rm
QCD}^2) &=& Z_q^{\mathrm {latt}}(a^2\,{p}^2, a^4p^{[4]}, a^6 p^{[6]}, ap_4,
 a^2\Lambda_{\rm
QCD}^2)\no &-&  
c_{a2p4} \, a^2 \frac{p^{[4]}}{p^2} - c_{a4p4}\,a^4\,p^{[4]}
\eea  
The result is shown in the r.h.s of \fig{fig:NP_subtracted}, one point per cubic
orbit. The non-perturbatively corrected 
$Z_q^{\mathrm {hyp\_corrected}}(a^2p^2,a^2\Lambda_{\rm QCD}^2)$ of 
\eq{eq:owfexpan} is represented by the black line in the r.h.s of 
\fig{fig:NP_subtracted}. It is well in the middle of the subtracted points, as
we would expect.

It is seen that the half-fishbones have been strongly reduced. One
sees a remainder of these artefacts due to the less democratic, or
``tyrannic'' points. These have only one non-vanishing component or one
large and a very small one. These points are not so many as seen in the plot,
their orbits are small which explains the larger error bars. 

We have checked that these tyrannic points have indeed a small impact on the
hypercubic corrected result.

\subsubsection{The slopes in $p^{[4]}$}

\begin{figure}[hbt]
  \begin{center}
    \includegraphics[width=100mm]{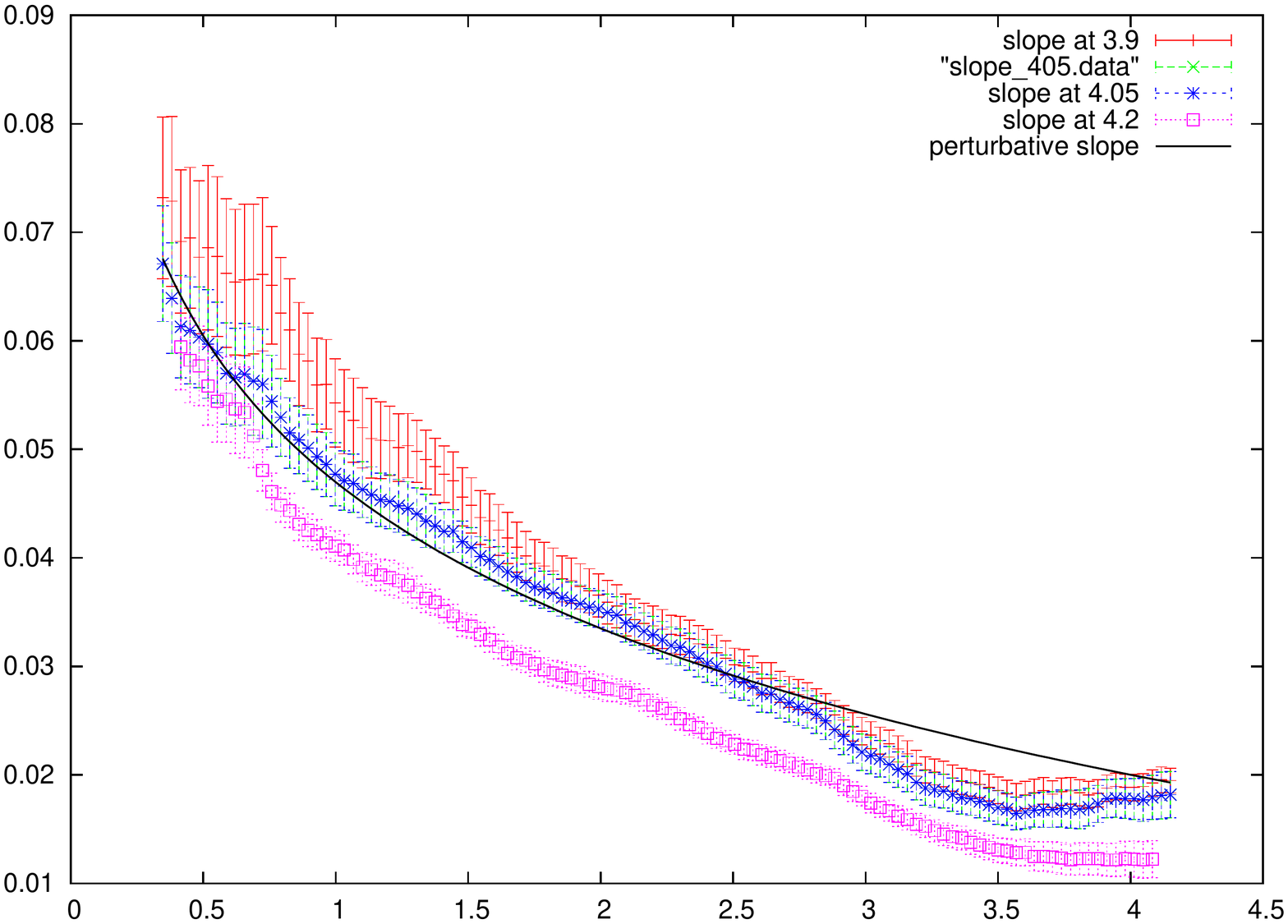}
  \end{center}
\caption{\small The three lattice slopes $R$ defined in \eq{eq:deriv} for the three lattice
 spacings and  the perturbative one 
$R_{\mathrm{pert}}=c_{q3}'+ c_{q4}\log(a^2p^2)$  for $\beta$ = 3.9, 
in terms of $a^2 p^2$ in the horizontal axis.}
\label{fig:slopes}
\end{figure}

The sliding window fit solves for every window a value for the slope 
$R(a^2p^2)$~\eq{eq:deriv}, i.e. the derivative  $\partial Z_q^{\mathrm
{latt}}/\partial (a^2p^{[4]}/p^2)$. This allows for a study of the shape of this
function $R$. In \fig{fig:slopes} we plot this slope $R$ defined in
\eq{eq:deriv} for the   three values of $\beta$. We also plot the equivalent
slope using the perturbative formula with the $p_\mu$ prescription for
$\beta=3.9$,  section~\ref{sec:subpert}: $R_{\mathrm{pert}}$ = $c_{q3}' + c_{q4}\log(a^2p^2)$,
$c_{q3}'$ defined  in \eq{eq:feli2} and $c_{q4}$ in \eq{eq:feli}. We see that 
this perturbative slope is in fair agreement with the non-perturbative one,
explaining the good elimination of half-fishbones in the r.h.s. in
\fig{fig:pert}.

 The three non-perturbative data in \fig{fig:slopes} give the impression to be 
 affine (a constant minus a linear term) over a rather large momentum interval. This is what is expressed in
 \eq{eq:ow}  from where we have deduced the one-window fit: a fit over the full
  range [0.5-3.5] with  two hypercubic parameters only~\footnote{Of course there
  are additionally as many hypercubic insensitive parameters as there are 
  values of $p^2$ in the range, which are simply the values of
   $Z_q^{\mathrm {hyp\_corrected}}(a^2p^2,a^2\Lambda_{\rm QCD}^2)$.}

\begin{figure}[hbt]
\begin{center}
\begin{tabular}{cc}
\includegraphics[width=8.5cm]{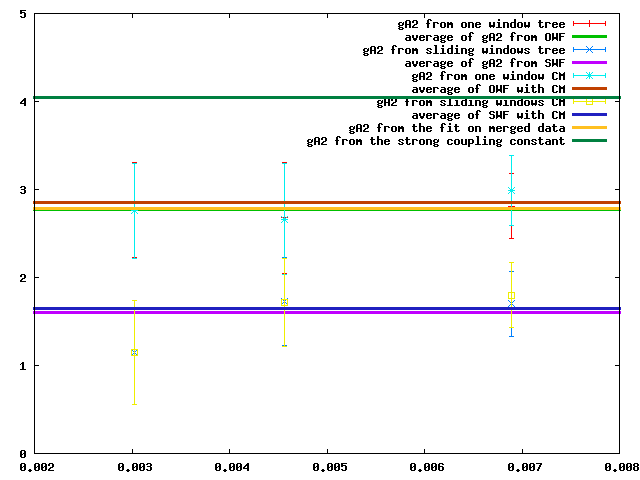} &
\includegraphics[width=8.5cm]{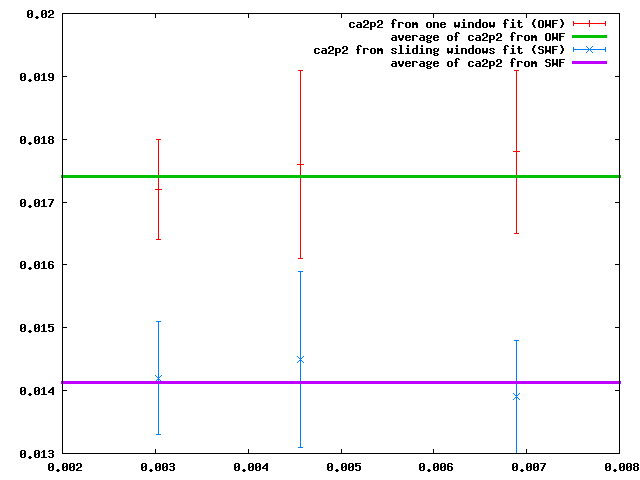}
\end{tabular}
\end{center}
\caption{\small We plot the values of the fitted slopes $c_{a2p2}$ (r.h.s) and the
condensate $g^2\, \VEV{A^2}$ (l.h.s) as extracted from the $1/p^2$ contribution
to the fit, see table~\ref{tab:p2andA2}, \ref{tab:p2andA2_sliding} and
\ref{tab:merged}. In the
left plot we show the results from the OWF and from the SWF. It can be seen that
the $O(\alpha^4)$ formula for the Wilson coefficient computed by Chetyrkin and 
Maier~\cite{Chetyrkin:2009kh} of $\VEV{A^2}$ (indicated by the "CM" initials)
is about 20 \% below the tree level result.
We show the value obtained from the merged results of the three  lattice
spacings, table~\ref{tab:merged}. Finally, for the sake of comparison we show the result
from the strong coupling constant of~\cite{Blossier:2010ky}. The horizontal axis
is $a^2$ in fm$^2$.}
\label{fig:p2andA2}
\end{figure}
\begin{figure}[hbt]
  \begin{center}
    \includegraphics[width=100mm]{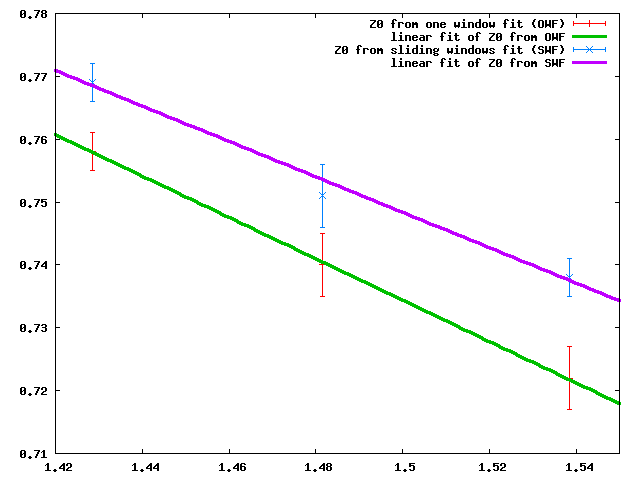}
  \end{center}
\caption{\small We show the value of $Z_q^{\mathrm{pert}}$ defined in \eq{eq:zqfit} for all 
three lattice spacings as a function of the bare coupling constant 
$g^2= 6/\beta$.}
\label{fig:g2Z0}
\end{figure}
\begin{table}[h]
\centering
\begin{tabular}{||c|c|c|c|c|c|c||}
\hline
\hline
$\beta$ & $a^2$ fm$^2$ & $c_{a2p4}$ & $c_{a4p4}$ & $c_{a2p4}/g^2$ &
$c_{a4p4}/g^2$ & $\chi^2$/d.o.f
\\ \hline
$3.9$ & 0.00689 & 0.067(4) &-0.0149(10)&0.044(3)&-0.0097(7)& 4.1 \\ \hline
$4.05$ &0.00456& 0.065(3)& -0.0144(5) & 0.044(2)&-0.0097(3)&0.53\\ \hline
$4.2$ &0.00303&  0.055(11)& -0.0124(4)&0.039(8) &-0.0089(3)&0.98 \\ \hline
\end{tabular}
\caption{\small Results for the slope in $a^2 \,p^{[4]}/p^2$  and $a^4 \,p^{[4]}$ 
and the same divided by $g^2$ in the one window fit.}
\label{tab:p4slopes}
\end{table}

The fitted values for $c_{a2p4}$ and $c_{a4p4}$ from the one window fit are given in 
table~\ref{tab:p4slopes} as well as the same divided by $g^2$, since
perturbation theory expects at least for $c_{a2p4}$ to be $\propto g^2$.
Before dividing by $g^2$ a small scaling violation is apparent, which
corresponds to the non overlap of the curves at different $\beta$'s in  \fig{fig:slopes}
It appears on table~\ref{tab:p4slopes} that dividing by $g^2$ improves 
significantly the scaling. The $\chi^2$ in 
table~\ref{tab:p4slopes} is not good for $\beta=3.9$, apparently due to some
structure at the lower end of the plot.

\section{Running including $\VEV{A^2}$  corrections from OPE}
\label{sec:running}
\begin{table}[h]
\centering
\begin{tabular}{||c|c|c|c|c|c||}
\hline
\hline
$\beta$ & $a^2$ fm$^2$ & $Z_q^{\mathrm{pert}}$ & $c_{a2p2}$ & 
$g^2 \VEV{A^2}_{\mathrm tree }$ & $g^2 \VEV{A^2}_{\mathrm
CM}$ 
\\ \hline
$3.9$ & 0.00689 & 0.726(5) &0.0201(13)&  3.20(38)  & 2.62(31)   \\ \hline
$4.05$ &0.00456& 0.742(5) &0.0200(15)&  3.09(65)  &2.57(54)\\ \hline
$4.2$ &0.00303&  0.760(3)  & 0.0194(8) &3.23(55)  & 2.74(47) \\ \hline
average & & &  0.0201(3) &3.18(28) &  2.64(23) \\ \hline
\end{tabular}
\caption{\small Results for  $Z_q^{\mathrm{pert}}$(10GeV)  and 
$c_{a2p2}$~\eq{eq:zqfit} from the one-window-fit and the estimated $g^2
\VEV{A^2}$ vev from the $1/p^2$ term and  from the the
Chetyrkin-Maier~\cite{Chetyrkin:2009kh} (CM) Wilson coefficient. Notice that
$Z_q^{\mathrm{pert}}$ and $c_{a2p2}$  from these two fits are very close.}
\label{tab:p2andA2}
\end{table}

\begin{table}[h]
\centering
\begin{tabular}{||c|c|c|c|c|c||}
\hline
\hline
$\beta$ & $a^2$ fm$^2$ & $Z_q^{\mathrm{pert}}$ 
&$c_{a2p2}$ 
& $g^2 \VEV{A^2}_{\mathrm tree}$ &
 $g^2 \VEV{A^2}_{\mathrm CM}$ 
\\ \hline
$3.9$ & 0.00689 & 0.741(3)&0.0161(9)  &2.07(37)& 1.70(31)\\ \hline
$4.05$ &0.00456& 0.753(5) &0.0168(14) &2.13(52)& 1.78(43)\\ \hline
$4.2$ &0.00303& 0.771(3)  &0.0164(9)  &1.59(60)& 1.36(51) \\ \hline
average & & &    0.0165(6)  &1.99(26)(27) &  1.65(22)(27) \\ \hline
\end{tabular}
\caption{\small The same as in table~\ref{tab:p2andA2} using the data
from the sliding-windows-fit to hypercubic corrections.}
\label{tab:p2andA2_sliding}
\end{table}
In this section we will check the running of $Z_q$. 
To this purpose, we shall use both
the formula \eq{eq:app_final2} derived in the appendix~\ref{appendix2}, to which
we add a lattice artefact term $\propto a^2p^2$ not yet subtracted:
\bea\label{eq:zqfit}
Z_q^{\mathrm {hyp\_corrected}}(a^2p^2)
  &=& Z_q^{\rm pert\; RI'}(\mu'^2) \,
    c_{0 Z_q}^{\rm RI'}\left(\frac{p^ 2}{\mu'^2},\alpha(\mu')\right)  \nonumber \\
    &\times& \left( 1 \ + \,
    \frac{c_{2 Z_q}^{\mMSB}\left(\frac{p^ 2}{\mu^2},\alpha(\mu)\right)}
    {c_{0 Z_q}^{\rm RI'}\left(\frac{p^ 2}{\mu^2},\alpha(\mu)\right)} \
 \frac{c_{2 Z_q}^{\rm RI'}\left(\frac{p^ 2}{\mu^2},\alpha(\mu)\right)}
     {c_{2 Z_q}^{\mMSB}\left(\frac{p^ 2}{\mu^2},\alpha(\mu)\right)} \
    \frac{\ \langle A^2\rangle_{R,\mu^2}}{32 \ p^2}  \right) \nonumber \\
&+& c_{a2p2}\; a^2\,p^2 \ .
\eea
and a formula including only an OPE correction with a tree-level Wilson coefficient,
\bea\label{eq:zqfit-tl}
Z_q^{\mathrm {hyp\_corrected}}(a^2p^2) \ = \
Z_q^{\rm pert\; RI'}(\mu'^2) \,
 c_{0 Z_q}^{\rm RI'}\left(\frac{p^ 2}{\mu'^2},\alpha(\mu')\right)
 \ \left( 1 \ + \frac{c_{1overp2}} {p^2} \right) + 
c_{a2p2}\; a^2\,p^2 \ ,
\eea
where $c_{1overp2}=g^2\langle A^2 \rangle/12$. We use $\mu'=\mu = 10$ GeV as the renormalisation scale ; 
$c_{0Z_q}^{\rm RI'}(p^2,\mu^2)$ is
computed from the four loop perturbative running of  
$Z_q$~\cite{Chetyrkin:1999pq}:
\bea\label{eq:C0def}
c_{0Z_q}^{\rm RI'}(p^2,\mu^2) \equiv 
\frac{Z_q^{\mathrm{pert\; RI'}}(p^2, g^2_{\mathrm{bare}})}
{Z_q^{\mathrm{pert\; RI'}}(\mu^2, g^2_{\mathrm{bare}})} \ ;
\eea
 $c_{2Z_q}^\mMSB(p^2,\mu^2)$
is the three loop Wilson coefficient of $\VEV{A^2}$ in the expansion of
$Z_q$~\cite{Chetyrkin:2009kh} and the ratio
\beq
\frac{c_{2 Z_q}^{\rm RI'}\left(\frac{p^ 2}{\mu^2},\alpha(\mu)\right)}
{c_{2 Z_q}^{\mMSB}\left(\frac{p^ 2}{\mu^2},\alpha(\mu)\right)} \ = \
\frac{1 - 0.1317 \ \alpha^2(\mu) - 0.5155 \ \alpha^3(\mu) } {1 - 0.1317 \ \alpha^2(p) - 0.5155 \ \alpha^3(p) }
\eeq
was obtained in appendix A. We express the lattice spacing
(cut-off) dependence as a dependence in $g^2_{\mathrm{bare}}$.  
 $Z_q^{\mathrm{pert\; RI'}}(\mu^2, g^2_{\mathrm{bare}})$
is the perturbative contribution to $Z_q$ at the scale $\mu$ in the RI'-MOM
 scheme.  
In other words, 
\bea\label{eq:defrimom}
Z_q^{\mathrm {RI'}}(p^2,g^2_{\mathrm{bare}})&=& 
Z_q^{\mathrm{pert}\; RI'}(p^2, 
g^2_{\mathrm{bare}}) \nonumber \\
&\times& \left( 1 \ + \,
\frac{c_{2 Z_q}^{\mMSB}\left(\frac{p^ 2}{\mu^2},\alpha(\mu)\right)}
{c_{0 Z_q}^{\rm RI'}\left(\frac{p^ 2}{\mu^2},\alpha(\mu)\right)} \
\frac{c_{2 Z_q}^{\rm RI'}\left(\frac{p^ 2}{\mu^2},\alpha(\mu)\right)}
{c_{2 Z_q}^{\mMSB}\left(\frac{p^ 2}{\mu^2},\alpha(\mu)\right)} \
\frac{\ \langle A^2\rangle_{R,\mu^2}}{32 \ p^2}  \right)
\eea
\hfill\break

From now on  $Z_q^{\mathrm{pert}}$ will refer to $Z_q^{\mathrm{pert}\;
RI'}$.
We fit three parameters: $Z_q^{\mathrm{pert}}$, $c_{a2p2}$ and alternatively
$c_{1overp2}$ (which amounts to a tree level treatment of the $c_{2Z_q}$
coefficient) or the vev $g^2\VEV{A^2}$.  In order to estimate the systematic
errors we will treat in parallel the  one window fit and the sliding window
one.  The results are reported in table~\ref{tab:p2andA2}, 
table~\ref{tab:p2andA2_sliding} ,  \fig{fig:p2andA2} and \fig{fig:g2Z0}. The
coefficient $c_{a2p2}$ obviously refers to  an $O(4)$ invariant lattice spacing
artefact which is not detected by our non-perturbative hypercubic correction
method. We see in the right plot  of fig.~\ref{fig:p2andA2} as well as in the
tables that this coefficient scales very well when expressed in lattice units,
as it should be. The  coefficient of $1/p^2$, if it is related to a vev  $
\VEV{A^2}$, should rather scale in physical units. We see in the left plot of
fig.~\ref{fig:p2andA2} that a constant value is rather well verified although
with large errors.  The results presented in table~\ref{tab:p2andA2} and
\ref{tab:p2andA2_sliding} show that the estimates for $g^2\VEV{A^2}$ from OPE expressions 
with Wilson coefficient from the Chetyrkin-Maier (CM) three-loop expression 
are systematically about 20\% below the ones from the tree level one.


\subsection{Analysis from the non-perturbative hypercubic corrections}
\subsubsection{Comparison of the running from the OWF and the SWF}
From tables \ref{tab:p2andA2} and~\ref{tab:p2andA2_sliding} we see 
that $g^2\,\VEV{A^2}$ is systematically larger for the OWF than for the SWF.
At first sight it seems surprising since the OWF and SWF hypercubic corrected
data are very similar, see \fig{fig:NP_subtracted}. One reason is the
correlation between $c_{a2p2}$ and $g^2\,\VEV{A^2}$:
 $c_{a2p2}$ is also systematically larger for the OWF 
than for the SWF. This correlation is understandable as the $a^2p^2$ increases
with $p^2$ while $1/p^2$ decreases. This is compensated by a $Z_q^{\mathrm{pert}}$ 
smaller for the OWF than for the SWF. We will consider these differences as a systematic
uncertainty in our fits and count them in the errors. 
 
\subsubsection{Dependence on the fitting range}

\begin{figure}[hbt]
  \begin{center}
  \begin{tabular}{cc}
    \includegraphics[width=85mm]{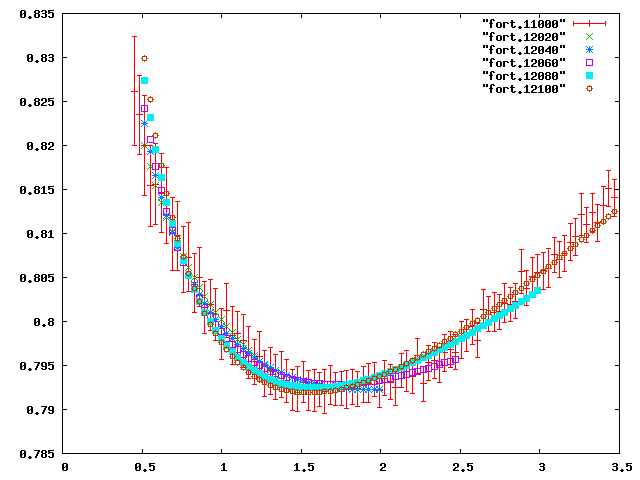}
 \includegraphics[width=85mm]{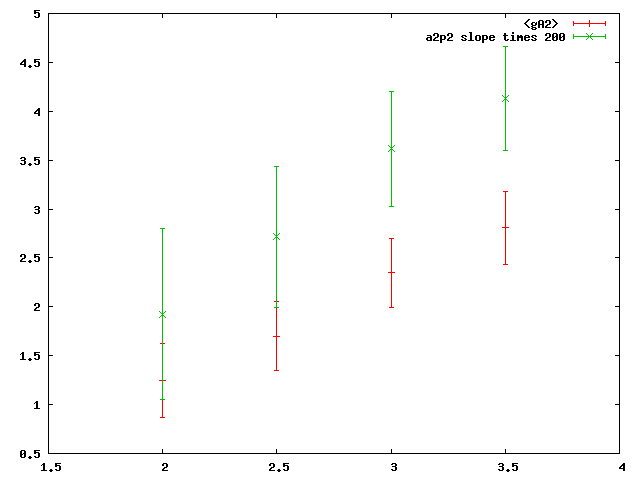}
	\end{tabular}
  \end{center}
\caption{\small In the r.h.s plot we show how $g^2\VEV{A^2}$, fitted with the CM 
Wilson coefficient, depends on the upper
range of our fits, always starting at $a^2p^2=0.5$ for $\beta=3.9$. The points
correspond to $a^2p^2 < 2.0, 2.5, 3.0, 3.5$. The r.h.s plot also shows the
$a^2p^2$ artefact slope. We find again a positive correlation between both series
of data. The l.h.s of the plot illustrates this, showing for the same data how the
fitting function depends on the upper bound of the range. }
\label{fig:range}
\end{figure}

\begin{table}[h]
\centering
\begin{tabular}{||c|c|c|c|c||}
\hline
\hline
Upper bound & $Z_q^{\mathrm{pert}}$ & $c_{a2p2}$ & 
$g^2 \VEV{A^2}_{\mathrm tree }$ & 
$g^2 \VEV{A^2}_{\mathrm CM }$
\\ \hline
2.0 &  0.754(6) &0.0089(21)&  1.58(39)  &1.28(32)   \\ \hline
2.5 & 0.745(6) & 0.0130(18)& 2.05(37) &  1.67(30)   \\ \hline
3.0 &  0.733(5)  & 0.0175(15)  & 2.73(36) & 2.22(30)  \\ \hline
3.5 &  0.726(5)  &  0.0201(13) & 3.20(38) &  2.62(31) \\ \hline
\end{tabular}
\caption{\small $\beta=3.9$: results for the  $Z_q^{\mathrm{pert}}$ ($\mu$= 10\, GeV)  
and $c_{a2p2}$~\eq{eq:zqfit} from the one-window hypercubic corrected 
data and the estimated $g^2 \VEV{A^2}$ vev from the $1/p^2$, plotted as a
function of the upper bound of the fitting range (in GeV).}
\label{tab:range}
\end{table}
An additional test is to look for the effect of the fitting range. The
results are shown in table~\ref{tab:range} and \fig{fig:range}. One sees again a
correlation between the $c_{a2p2}$ slope and $g^2\,\VEV{A^2}$. Both decrease when
the fitting range shortens while correlatively $Z_q^{\mathrm{pert}}$ increases. The l.h.s of
\fig{fig:range} explains how this happens: when the range is shorter, the error
bars allow for a less curved fit. But never the fit reaches a value such that
$g^2\,\VEV{A^2}$ disappears.  The shortest window $0.5 <a^2p^2 < 2.0$ gives
the smallest value for $g^2\,\VEV{A^2}$ but still 4 sigmas away from 0.  

\subsection{Analysis from the perturbative hypercubic corrections}
It is then useful to check if similar results are obtained after
a perturbative correction to the hypercubic artefacts has been applied.
\begin{table}[h]
\centering
\begin{tabular}{||c|c|c|c|c||}
\hline
\hline
prescription & $Z_q^{\mathrm{pert}}$ & $c_{a2p2}$ & 
$g^2 \VEV{A^2}_{\mathrm tree }$ & 
$g^2 \VEV{A^2}_{\mathrm CM }$ 
\\ \hline
$\tilde p_\mu$  & 0.712(11) &  -0.0026(23)&  2.98(1.49) & 2.53(1.23) \\ \hline
 $p_\mu$ & 0.745(3) & -0.0061(8)&  1.76(35)&  1.45(29)     \\ \hline
\end{tabular}
\caption{\small For $\beta=3.9$, results for the $Z_q^{\mathrm{pert}}(\mu^2, 
g^2_{\mathrm{bare}})$
($\mu$= 10\, GeV)  and $c_{a2p2}$~\eq{eq:zqfit}
and $g^2 \VEV{A^2}$ from the lattice data after a perturbative  
hypercubic correction.  $g^2 \VEV{A^2}$ is estimated at tree level from
the $1/p^2$ contribution.}
\label{tab:pert_cond}
\end{table}
We have used two prescriptions to apply the perturbative corrections. With the
data obtained from the $\tilde p_\mu$ prescription,
section~\ref{sec:subpertild}, we perform an average on all the cubic orbits of
every $p^2$ after a democratic selection $p^{[4]}/(p^2)^2 < 0.3$. This leaves us
with not too many points and it results in rather large statistical errors.
We then perform the same running fit than on the non-perturbatively hypercubic
corrected results: we fit with one perturbative running contribution, one
$1/p^2$ contribution and one $\propto a^2p^2$ artefact. 
With the data from the $p_\mu$ prescription, section~\ref{sec:subpert}, we
perform the same fit over an average on all the cubic orbits of every $p^2$
without any democratic selection, since the hypercubic artefacts have already 
been efficiently reduced. The results are in table~\ref{tab:pert_cond}. 

A first remark is that the $c_{a2p2}$ coefficients are compatible with zero,
indicating that the perturbative correction has efficiently eliminated 
this artefact. 
The coefficient of the $1/p^2$ non-perturbative contribution is found different 
from zero, in the same ballpark as the results from the non-perturbative
hypercubic correction, in tables~\ref{tab:p2andA2} and~\ref{tab:p2andA2_sliding}. 
 The $\tilde p_\mu$ prescription has too large errors to be conclusive
but the $p_\mu$ is five sigmas away from zero, very similar to 
the results in table~\ref{tab:p2andA2_sliding}.
The value of $Z_q^{\mathrm{pert}}$ for the  $\tilde p_\mu$
prescription is rather low but compatible within less than two sigmas from the
result for $\beta=3.9$ in  table~\ref{tab:p2andA2_sliding}.
 
\subsection{Running of $Z_q^{\rm pert}$}
It is interesting to consider the dependence of $Z_q^{\mathrm{pert}}$ as a
function of  $g^2$. This is plotted in fig.~\ref{fig:g2Z0} both for the OWF and
the SWF. It is strikingly linear, specially for the OWF. Indeed 
from~\cite{Constantinou:2009tr}, eq.~(24) perturbation theory gives a linear
dependence with a slope $\simeq -0.19$. This comes from the coefficient $b_{q1}$
in  \eq{eq:pert} which is multiplied by $g^2$. In our case we find from the OWF
\bea\label{eq:zqone}
Z_q^{\rm pert}((10\, \mathrm {GeV})^2,g^2_{\mathrm {bare}})&=& 
0.737(3) - 0.313 (6) \,(g^2_{\mathrm {bare}}-1.5) \ , \no 
Z_q^{\rm pert}((2\, \mathrm {GeV})^2,g^2_{\mathrm {bare}})&=& 0.766(3) - 
0.324(6) \,(g^2_{\mathrm {bare}}-1.5) \ ;
\eea
 and from the SWF
\bea\label{eq:zqmany}
Z_q^{\rm pert}((10 \mathrm {GeV})^2,g^2_{\mathrm {bare}}) &=& 
0.751(2)(7) - 0.273(6)\left(^{0.002}_{-0.038}\right) \,(g^2_{\mathrm {bare}}-1.5) \ , 
\no 
Z_q^{\rm pert}((2\, \mathrm {GeV})^2,g^2_{\mathrm {bare}}) &=& 
0.780(3)(7) - 0.284(6) \,\left(^{0.002}_{-0.040}\right)(g^2_{\mathrm {bare}}-1.5) \ .
\eea
We see that the coefficients of $g^2$, 
 \bea
  \frac {\partial Z_q^{\mathrm{pert}}((2\, \mathrm {GeV})^2,g^2_{\mathrm {bare}})}
  {\partial g^2} = \left\{ 
\begin{array}{lr} 
-0.329(6) & \mathrm{OWF} \\ 
-0.287(26) & \mathrm{SWF}
\end{array}
\right. \ ,
 \eea
 are significantly larger than the perturbatively expected, -0.19. But the linear 
 behaviour predicted by perturbation theory is well verified, especially 
 for OWF.

\begin{figure}[hbt]
  \begin{center}\begin{tabular}{cc}
    \includegraphics[width=85mm]{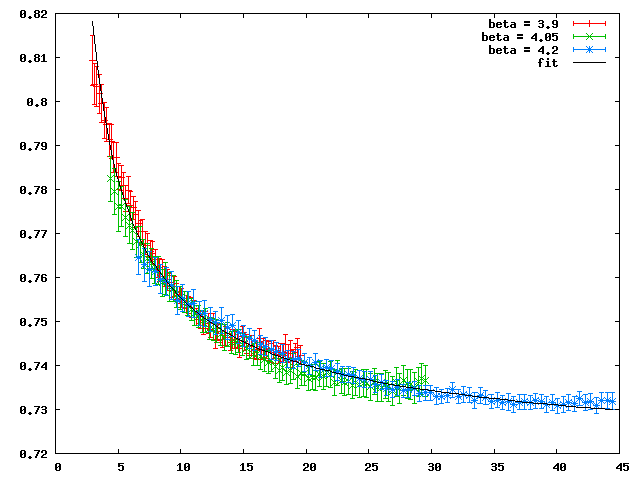}&
      \includegraphics[width=85mm]{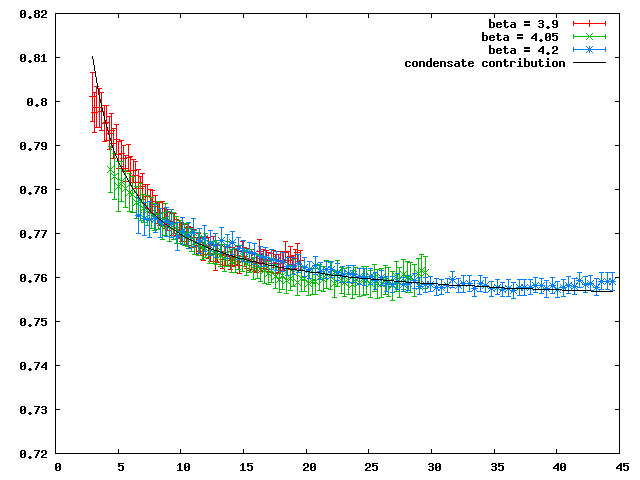}
\end{tabular} \end{center}
\caption{\small The merged plot with the OWF results at $\beta=4.05$ and
$\beta=4.2$ rescaled to the $\beta=3.9$ thanks to the ratios of 
$Z_q^{\mathrm{pert}}$ given in table~\ref{tab:p2andA2}.  The l.h.s shows  the
data corrected for all lattice artefacts. The r.h.s shows the same data
furthermore corrected by the perturbative  running factor up to 10 GeV. The
horizontal axix is $p^2$ in GeV $^2$. The black line on the l.h.s corresponds to
the global fit with  perturbative running and CM (three loops) wilson
coefficient for the  $1/p^2$ term.  The black line on the r.h.s corresponds only
to the $1/p^2$ times the three loops wilson coefficient added to  $Z_q^{\mathrm
{pert}}((10\, \mathrm {GeV})^2, 6/3.9)=0.726$}
\label{fig:all_betas}
\end{figure}
\subsection{Merging the three lattice spacings.}

From \eq{eq:zqfit} and \eq{eq:defrimom} it is clear that
\bea\label{eq:artefree}
Z_q^{\mathrm {RI'}}(p^2,g^2_{\mathrm{bare}})=
Z_q^{\mathrm {hyp\_corrected}}(a^2p^2) - c_{a2p2} a^2p^2 \ .
\eea
In this section we use the one window fit, section~\ref{sec:one}, and
 the momentum $p^2$ is now expressed in physical units. For the coefficient 
$c_{a2p2}$ we use the values in table~\ref{tab:p2andA2}:
$c_{a2p2}=0.0174$. The three $Z_q^{\mathrm {RI'}}$ for the three
$\beta$'s do not match due to the running of 
$Z_q$ as a function of the lattice spacing.  
To make them match it turns out that it is enough to take into account 
the ratios of $Z_q^{\mathrm{pert}}$'s given in table~\ref{tab:p2andA2}.  
We plot on the l.h.s of fig.~\ref{fig:all_betas} the three sets of data  where
 the $\beta=4.05, 4.2$ ones have been rescaled to the $\beta=3.9$
 scale.
We see a rather good overlap.  There is however a
flattening at the right end of every $\beta$ which stays within one sigma
from the other $\beta$'s. 
We understand it as a failure of the hypercubic artefacts treatment. 
On the r.h.s side of fig.~\ref{fig:all_betas} we plot the same number 
corrected for perturbative running by the multiplicative factor  
$0.726/Z_q^{\rm pert}(p^2, 6/3.9)$, where 0.726 is taken from 
table~\ref{tab:merged}. The black line is just the non-perturbative
contribution added to 0.726.
The comparison of both plots in \fig{fig:all_betas} is enlightening.
We see that the non perturbative term contributes about one half of the change 
between the smallest momenta and the largest ones. Both the perturbative running
and the non perturbative contribution are convex, which makes it
difficult to disentangle them.  But we also see that the
perturbative running cannot account for the full variation of the data.  
The best-fit parameters resulting from this merged analysis can be found 
in table~\ref{tab:merged}, where we use $g^2_{\mathrm{bare}}=6.0/3.9$ since we have rescaled
all the data to the $\beta=3.9$ one. 
The values in table~\ref{tab:merged} for $g^2(\mu^2) 
\VEV{A^2}_{\mu^2\;\mathrm{merged}}$
turn out to be rather central in the set of values of 
tables~\ref{tab:p2andA2} and~\ref{tab:p2andA2_sliding}.  
The value for $Z_q^{\mathrm{pert}}(10 \,{\mathrm GeV}, 6./3.9)$ is in very good agreement
with \eq{eq:zqone}: $Z_q^{\mathrm{pert}}(10 \,{\mathrm GeV}, 6./3.9)=0.725(3)$.

\begin{table}[h]
\centering
\begin{tabular}{||c|c|c||}
\hline
\hline
  $Z_q^{\mathrm{pert}}$ &  
$g^2 \VEV{A^2}_{\mathrm tree }$ & 
$g^2 \VEV{A^2}_{\mathrm CM }$
\\ \hline
 0.726(2) &  3.13(43)  &2.55(36)   \\ \hline
\end{tabular}
\caption{\small Merged data from three $\beta$'s: results for the  $Z_q^{\mathrm{pert}}$ 
($\mu$= 10\, GeV) rescaled to $\beta=3.9$, from the one-window hypercubic
 corrected data (OWF) with tree level and the three loop 
 formula, \eq{eq:app_final2}.}
\label{tab:merged}
\end{table}

\subsection{Summarizing}

Many of our results for $g^2 \VEV{A^2}$ are shown in the l.h.s of 
\fig{fig:p2andA2}.
We did not plot the range dependent data to not overload the plot but they fall
in the range covered by the data plotted in~\fig{fig:p2andA2}.
To summarise our results and estimate the systematic uncertainty
we consider the set of values in 
tables~\ref{tab:p2andA2}, \ref{tab:p2andA2_sliding},\ref{tab:range} and
\ref{tab:merged}. We will make a separate average for the tree level data and
the $O(\alpha^4)$ (CM) ones since the comparison with estimates from other 
quantities, such as the coupling constant, need to be performed in the 
same scheme, expansion, order and scale. The scheme is $\MSB$ and the precise
implementation is detailed in the appendix. We computed an average of all
above-listed data, weighted by their inverse squared error. The inverse squared 
statstical error is  the sum of the inverse squared errors. The systematic error
is taken such as to incorporate the central values within the error-bars.
This is rather conservative. 
For $Z_q^{\rm pert}$ we average \eq{eq:zqone} and \eq{eq:zqmany}
with a similar method. We get:
\beq\label{eq:final}
g^2(\mu^2) \VEV{A^2}_{\mu^2\; tree} &=& 2.45(14)\left(^{+0.78}_{- 0.87}\right)
 \;\mathrm {GeV}^2  \quad \mu=10\,
\mathrm{GeV} \ ,
\no
g^2(\mu^2) \VEV{A^2}_{\mu^2\; CM} &=& 2.01(11)\left(^{+0.61}_{- 0.73}\right) 
 \;\mathrm {GeV}^2  \quad \mu=10\,
\mathrm{GeV} \ ,
\no
Z_q^{\rm pert}((10\,{\mathrm {GeV}})^2,g^2_{\mathrm {bare}}) 
&=& 0.744(2)(7) - 0.311(6)\left(^{+0.002}_{- 0.038}\right) 
\,(g^2_{\mathrm {bare}}-1.5) \ ,
\no
Z_q^{\rm pert}((2\,{\mathrm {GeV}})^2,g^2_{\mathrm {bare}}) 
&=& 0.773(3)(7) - 0.323(6)\left(^{+0.002}_{- 0.040}\right) 
\,(g^2_{\mathrm {bare}}-1.5) \ ;
\eeq
where the first error is statistical and the second the systematic one.
The values of $Z_q^{\mathrm {RI'}}(p^2,g^2_{\mathrm{bare}})$ may then
be derived from \eq{eq:defrimom}:

\beq\label{eq:finalbis}
Z_q^{\mathrm {RI'}}((2\,{\mathrm {GeV}})^2,g^2_{\mathrm {bare}}) 
&=& 0.805(14) - 0.336(6) \left(^{+0.002}_{- 0.042}\right)
\,(g^2_{\mathrm {bare}}-1.5) \ . \qquad {\rm [CM]} 
\eeq
The results obtained from the perturbative hypercubic correction, 
table~\ref{tab:pert_cond} have not been used in the final estimate (because
we did not compute them for all $\beta's$) but they fall
within the bounds at less than one sigma.  
 
Finally two lines in the l.h.s. plot of fig.~\ref{fig:p2andA2} show the
results for $g^2 \VEV{A^2}$ obtained from those in tab.~3 of ref.~\cite{Blossier:2010ky}:
\bea\label{eq:fromalpha}
g^2 \VEV{A^2}_{10 \;{\mathrm {GeV}}} = 
\left\{ \begin{array}{lr}
4.1 \pm 1.5 \ {\rm GeV}^2 & {\rm leading~log}\\
2.5 \pm 0.9 \ {\rm GeV}^2 & {\cal O}\left(\alpha^4\right) \\
\end{array}
\right. \ .
\eea
These values in \eq{eq:fromalpha} come from a totally different quantity: they
have been extracted from the running of the ghost-gluon coupling constant. 
The results of ref.~\cite{Blossier:2010ky} were obtained by applying an OPE formula 
including a Wilson coefficient approximated at the leading logarithm and at the 
order ${\cal O}\left(\alpha^4\right)$, but expanded in terms of $\alpha_T$. Then, 
in order to be properly compared with the results of this work, the values 
of \eq{eq:fromalpha} incorporate the correction by the effect of expanding the OPE formula 
in terms of the running coupling in $\overline{\rm MS}$. The lattice spacing applied 
in ref.~\cite{Blossier:2010ky} to get a physical scale, $a(3.9)=0.0801$ fm, was also 
slightly smaller than the one used in this work (see, for instance, 
tab.~\ref{setup}) and this has been also taken into account in 
obtaining \eq{eq:fromalpha}.

As a matter of fact, \eq{eq:fromalpha} exhibits a slower convergence of the perturbative  series of
the Wilson-coefficient as in the present paper. From tables~\ref{tab:p2andA2}, 
\ref{tab:p2andA2_sliding}, \ref{tab:range} and \ref{tab:merged} we see that  the
$O(\alpha^4)$ estimate is about 20 \% below the tree level while 
in table 3 of~\cite{Blossier:2010ky} it is about 45 \% below the leading 
logarithm one. The two estimates agree rather well  within 
the present accuracy.

\subsection{Conversion to $\overline{\mathrm {MS}}$}
The conversion to $\overline{\mathrm {MS}}$ can also be performed
from  $Z_q^{\mathrm{pert}}$ or $Z_q^{\mathrm {RI'}}$ which contains 
a non-perturbative contribution. Usually, in the literature, the values
are assumed to be perturbative. 
The conversion of $Z_q^{\mathrm{pert}}$ into $\overline{\mathrm {MS}}$ 
 will use the standard perturbative conversion formulae ~\cite{Chetyrkin:1999pq}. 
We get 
\beq
Z_q^{\overline{\mathrm {MS}}\,\mathrm{pert}}( (2\,{\mathrm {GeV}})^2, 
g^2_{\mathrm{bare}})/
Z_q^{\mathrm{pert}}((2\,{\mathrm {GeV}})^2, g^2_{\mathrm{bare}}) = 0.97 \ ,
\no
Z_q^{\overline{\mathrm {MS}}\,\mathrm{pert}}((2\,{\mathrm {GeV}})^2,
g^2_{\mathrm {bare}}) = 0.750(3)(7) - 0.313(20) 
\,(g^2_{\mathrm {bare}}-1.5) \ .
\label{eq:pertMS}
\eeq 
The central value of this result is about a 2 \% systematically below the 
results of~\cite{Constantinou:2010gr}. This can be presumably interpreted as 
a small systematic correction due to our subtraction of the non-perturbative contribution in obtaining 
$Z_q^{\rm pert}$ in the RI'-MOM scheme and converting it to $Z_q^{\mMSB\, \mathrm{pert}}$ at 
2 GeV.
 
From \eq{eq:app_final2} it is easy to see how to include the 
$g^2 \VEV{A^2}$ non perturbative contribution. Up to now we have applied 
the result of the appendix using RI'-MOM for the perturbative part and 
$\overline{\mathrm {MS}}$ for the ratio in the corrective parenthesis.
Had we wished to use the  $\overline{\mathrm {MS}}$ scheme for the perturbative 
contribution, we would have the main inconvenience of not to know the OPE 
contribution for $Z_q$ defined in the $\mMSB$ scheme. However, only in the aim 
of roughly estimating the non-perturbative correction, we can assume the 
OPE corrective parenthesis to remain the same and then get
\beq\label{eq:npMS}
Z_q^{\overline{\mathrm {MS}}\,\mathrm{non-perturbative}}((2\,{\mathrm {GeV}})^2,
g^2_{\mathrm {bare}}) = 0.781(6)(21) - 0.326(21) \,(g^2_{\mathrm {bare}}-1.5) \ .
\eeq
Notice that the non perturbative contribution is about 4\% at 2 GeV. 
Nevertheless, the results of~\cite{Constantinou:2010gr} have been 
obtained at momenta larger than 2 GeV. Although their estimates and 
our $Z_q^{\rm RI'}$ may agree with each other, a subtraction of 
the non-perturbative contribution in obtaining $Z_q^{\rm pert}$, 
probably still required at the fitting window of~\cite{Constantinou:2010gr} 
but not applied, could explain the small discrepancy of about a 2 \% that 
we discussed above. The discrepancy is anyhow only affecting the conversion 
of $Z_q^{\rm RI'}$ to $Z_q^{\mMSB\, \mathrm{pert}}$ at 2 GeV.
In Table 6 of~\cite{Constantinou:2010gr} one finds results for 
$Z_q^{\overline{\mathrm {MS}}}$ obtained at 2 GeV by applying the standard perturbative conversion 
formulae~\cite{Chetyrkin:1999pq}. 
Indeed, they turn out to be in between our results 
in \eq{eq:pertMS} and \eq{eq:npMS}. Quoting in order \eq{eq:pertMS}, ref.~\cite{Constantinou:2010gr} 
and \eq{eq:npMS} we get: 0.738, 0.751 and 0.769 for $\beta=3.9$ and 0.755, 0.780 and 0.786 
for $\beta=4.05$. 

\subsection{Comparison of different estimates for the gluon condensate}

Let us now compare the present estimate of  $g^2 \VEV{A^2}$ to previous ones
at $N_f=2$ and $N_f=0$, all taken at the renormalisation scale of 10 GeV and, 
when needed, transformed to the very precise renormalization scheme for 
the OPE expansion defined in the appendix \ref{appendix2}.

In~\cite{Boucaud:2005rm} a quenched study of $Z_q$ using Wilson-Clover and
overlap fermions ended with values of $\VEV{A^2}_{\rm MOM}$ in 
the range 2.67-3.2 GeV$^2$ with typical errors 
of 0.3 GeV$^2$. Notice that this computation was performed only up to leading 
logarithm for the Wilson coefficient and that the choice was to expand the perturbative 
series in terms of the coupling renormalized in the MOM scheme (this is why we use for 
the VEV the label MOM). Then, we can apply the expressions derived in the appendix 
\ref{appendix2} to obtain the estimates for $g^2 \VEV{A^2}$ in the above-mentioned 
renormalization scheme appearing in table~\ref{tab:global_comp}. 
However it is advocated in~\cite{Boucaud:2005rm} that
the $1/p^2$ contribution only increases by 10\%  when going from MOM to $\mMSB$.
On the other hand we have seen that between tree level and three loops a
decrease of 20\% was observed. In~\cite{Boucaud:2005rm} an artefact 
$\propto a/p^2$ was observed. We do not see it in the present analysis 
since the scaling of $g^2 \VEV{A^2}$ as a function of the lattice spacing 
indicates no visible $1/p^2$ contribution dependent on $a$. 

In~\cite{Boucaud:2008gn} a summary was performed of different 
estimates of $g^2 \VEV{A^2}$ from gluonic quantities at $N_f=0$~:
$\alpha_s$ from the three gluon vertex with equal momenta on the three legs (symmetric) and
from the three gluon vertex with one vanishing momentum (asymmetric), the ratio between the 
ghost and gluon propagators and $\alpha_s$ from the ghost and gluon propagators,
using Taylor's theorem. The ones involving gluon and ghost propagators 
agree fairly well but the latter is the most accurate. It gives  
$g_T^2 \VEV{A^2} = 5.1^{+0.7}_{-1.1}$, although the applied OPE formula was obtained by 
expanding the involved perturbative series in terms of $\alpha_T$. After the 
appropriate transformation, one obtains the result shown in table~\ref{tab:global_comp}. 
We also quote in the table the estimate of $g^2 \VEV{A^2}$ from the symmetric 
three gluon vertex, more precise than the one coming from the asymmetric vertex, and that 
appeared to be much higher than the estimate from $\alpha_T$ and compatible with that from 
quark propagtor. In the case of the three-gluon estimates, no available 
$O(\alpha^4)$ Wilson coefficient can help us to go beyond the leading logarithm 
approximation. However, either comparing the leading-logarithm estimates of the 
ones approximated at the order $O(\alpha^4)$, a clear discrepancy (by a factor of about two) 
appears between the estimates from ghost and gluon propagators and those from vertices or 
the quark propagator. This could imply that some systematic uncertainty is not completely 
under control. One might, for instance, guess that $1/p^4$-contributions can be invoked to 
reduce that discrepancy. For this to happen, the $1/p^4$-contributions had to be negative, and 
to tend to increase the estimate of $g^2 \VEV{A^2}$, for the OPE formula of $\alpha_T$; while it 
had to be positive, and reduce $g^2 \VEV{A^2}$, for the quark propagator. Indeed, although no 
stable fit including $1/p^4$-contributions can be performed, the sign seems to be the right one 
for $\alpha_T$ in \cite{Blossier:2010ky}. Also the right sign of the contributions to 
$Z_q$ is found in ref.~\cite{Boucaud:2005rm}.

\begin{table}[h]
\centering
\begin{tabular}{|c|}
\hline
\rule[0cm]{4cm}{0cm} measurement (GeV$^2$)
\\
\begin{tabular}{|c|c|c|c|c|}
\hline
$N_f$ & order  $g^2 \VEV{A^2}$ & $Z_q$ & $\alpha_T$ & 3-gluon \\
\hline
0 & LL & 9.4(3) & 5.2(1.1) & 10(3) \\
\cline{2-5}
 &  $O(\alpha^4)$ & 9.0(3) & 3.7(8) & \\
\hline
2 & LL & 2.7(4) &  4.1(1.5) & \\
\cline{2-5}
 &  $O(\alpha^4)$ & 2.55(36) & 2.5(9)& \\
\hline
\end{tabular}
\rule[-1.3cm]{0cm}{1.3cm}
\\
\hline
\end{tabular}
\caption{\small Comparison of estimates of $g^2 \VEV{A^2}$ from different quantities 
an $N_f=0$ and $N_f=2$. All are taken at the scale $\mu= 10$\, GeV. 
LL means leading logarithm for the Wilson coefficient. $O(\alpha^4)$ refers to
Chetyrkine and Maier computation.  }
\label{tab:global_comp}
\end{table}


In~\cite{Blossier:2010ky} the strong coupling constant was computed along
similar lines to what is done here, on the same set of ETMC gauge
configurations with $N_f=2$.  The necessity of a non-perturbative 
$\propto 1/p^2$ contribution was also found and the resulting 
condensate, $g^2 \VEV{A^2}_{10 \;{\mathrm GeV}} = 2.3(8)$, obtained through an OPE formula 
approximated at the $O(\alpha^4)$-order and 
expanded in $\alpha_T$, can be properly transformed~\footnote{We have taken into account 
the different lattice spacing in~\cite{Blossier:2010ky}.} to give the value of 
table~\ref{tab:global_comp}, also quoted in \eq{eq:fromalpha}, which agrees 
strikingly well with the result 
of tab.~\ref{tab:merged} (the one we also quote in tab.~\ref{tab:global_comp}) 
or that of \eq{eq:final}. The value obtained through a leading-logarithm-approximated 
formula is also displayed in tab.~\ref{tab:global_comp}. 
In~\cite{Martinelli:1996pk} Martinelli and Sachrajda proposed a criterium to validate
the use of operator expansion which we apply in this paper. They concluded that one should
compare the difference of the highest order of the perturbative expansion for two different
quantities with the non-perturbative contribution and check that the former is small
compared to the latter.  We have compared the  highest order of the perturbative expansion
of $Z_q$ with the $1/p^2$ contribution and find that the ratio ranges between 1/10 and 1/3
depending on the momentum. This is a good indication of the validity of our use of the
operator expansion. Had we used the perturbative expansion only up to $O(\alpha)$ the
criterium would not have been fulfilled. 

Furthermore, all these estimates in table~\ref{tab:global_comp} show a clear tendency of 
a decrease of $g^2 \VEV{A^2}$ from $N_f=0$ to $N_f=2$. This might 
support an interpretation of $g^2 \VEV{A^2}$ as originating in 
instantons~\cite{Boucaud:2002nc} since the instanton density should decrease
with light dynamical masses.

\section{Conclusion}

We have studied with care the twisted quark propagators  produced on
the ETMC set of $N_f=2$ gauge configurations. Our goal was to concentrate on two
major issues: the correction for lattice spacing artefacts, particularly the
hypercubic ones, and the presence of a sizeable non perturbative contribution of the 
$A^2$ operator. The latter is expected to be sizeable because it was seen in the
quenched case~\cite{Boucaud:2005rm}, it was seen in the unquenched study of the 
strong coupling constant~\cite{Blossier:2010ky}   and since the Wilson
coefficient of $g^2 \VEV{A^2}$ is not small in $Z_q$~\cite{Chetyrkin:2009kh}.
This is an important issue since from our estimates it gives a  
$\sim 4 \%$  contribution at 2 GeV. A reliable estimate of this non-perturbative
correction needs a large enough fitting range, which allows to distinguish 
a $1/p^2$ contribution from perturbative logarithms. But the fitting window 
is restricted below by infrared effects and above by lattice spacing artefacts.
We thus need to improve our control on dominant lattice spacing artefacts which are of
two types: hypercubic ones and $\propto a^2p^2$ ones.

Concerning the hypercubic artefacts, we have summarized the non-perturbative
correcting method~\cite{Becirevic:1999uc,deSoto:2007ht} which we compared
systematically with the perturbative results of~\cite{Constantinou:2009tr}. 
$Z_q$ has very large hypercubic artefacts which display, as a function of $p^2$
 a ``half-fishbone'' very far from a smooth curve  (see \fig{fig:fishbone}).
We check carefully how these fishbones are ``swallowed'' by the corrective
methods. It is worthwhile to emphasize that the ``democratic'' method, prescribing 
for instance a cut on $p^{[4]}/p^2$ to drastically reduce the number of allowed 
hypercubic orbits, is not good enough to eliminate the fishbones and to leave us with a 
smooth curve for $Z_q$.

The perturbative method to correct hypercubic artefacts suffers from
some options left: what to take for the coupling constant,
use of $p_\mu$ or $\tilde p_\mu=a^{-1} \sin(a p_\mu)$ ?
We tried first  to stick to the prescription of~\cite{Constantinou:2009tr}
 and use the boosted coupling constant. This
reduces the hypercubic artefacts only up to $a^2p^2 \simeq 1.6$
(see \fig{fig:pert}, l.h.s).
Guided by the test on the fishbone reduction we then
propose a prescription based on the same perturbative formulae but using
$p_\mu$. For $Z_q$ this reduces the hypercubic artefacts up to 
 $a^2p^2 \simeq 3.5$ which has been our upper bound in this work 
 (see \fig{fig:pert}, r.h.s). 

We test also the non-perturbative method to correct hypercubic artefacts. We use
two prescriptions. The first one uses a sliding window and the second one uses
only one fitting window on the full momentum range. 

We find that the hypercubic artefacts are sufficiently well  described and cured
by two terms: $\propto a^2 p^{[4]}/p^2$ and $\propto a^4 p^{[4]}$. We fit the
coefficients of these quantities and check their scaling with $\beta$.

From the resulting hypercubic corrected function  $Z_q(a^2p^2,a^2\Lambda_{\rm
QCD}^2)$ we perform  fits which incorporate the perturbative running, a non
perturbative $1/p^2$ term (presumably related to $g^2 \VEV{A^2}$), and a
hypercubic insensitive lattice spacing artefact  proportional to $a^2p^2$. The
fits are good. The $a^2p^2$ term scales almost perfectly in lattice units, as
expected. The $g^2 \VEV{A^2}$ term scales rather well in physical units as
expected. The accuracy on $g^2 \VEV{A^2}$ is reduced by some correlations in the
fits: we see a correlation between the method used to correct hypercubic
artefacts and the estimated value of  $g^2 \VEV{A^2}$. We also see a correlation
between the fitting range and the resulting $g^2 \VEV{A^2}$. But  all values
of $g^2 \VEV{A^2}$ fall into the same ballpark and none of these fits 
can be done without such a positive contribution. To estimate the systematic
uncertainty we have considered a large panel of fitting methods. All at more
than four sigmas from zero except at $\beta=4.2$ with the sliding window
where they are only 2.5 sigmas above zero.  Comparing the fitted $\VEV{A^2}$ 
using the tree level Wilson coefficient and that using the three loops one, 
we find that the latter is about 20 \% below the former.

The perturbative contribution to $Z_q$, $Z_q^{\mathrm{pert}}$ has a linear
dependence in the bare lattice coupling: see \fig{fig:g2Z0} and \eq{eq:final},
as expected from perturbation theory, but with a larger coefficient, even when
the boosted coupling constant is used in perturbation theory.
  
We also merge all three $\beta$'s after having subtracted the $a^2p^2$ term
and rescaled the $\beta=4.05, 4.2$ to $3.9$ using the ratios of
$Z_q^{\mathrm{pert}}(\mu)$.  The overlap of the three data sets is rather good.
The need of a non-perturbative contribution is also visible there. 
Both perturbative and non-perturbative contributions decrease with the momentum
and are convex. This makes the separation difficult. Grossly speaking 
they share the decrease between 4 and 40 GeV$^2$ in equal parts.   

We have converted our results for $Z_q^{\mathrm {RI'}}$ and its perturbative
part $Z_q^{\mathrm{pert}}(\mu)$ into the $\mMSB$ scheme.
 
Combining all the results  we find, using the three loop Wilson coefficient: 
\beq\label{eq:resultA2}
g^2(\mu^2) \VEV{A^2}_{\mu^2\; CM} &=& 2.01(11)\left(^{+0.61}_{- 0.73}\right) 
 \;\mathrm {GeV}^2  \quad \mu=10\, \mathrm {GeV} \ ,
\no
Z_q^{\rm pert}((2\,{\mathrm {GeV}})^2,g^2_{\mathrm {bare}}) 
&=& 0.773(3)(7) - 0.323(6)\left(^{+0.002}_{- 0.040}\right) 
\,(g^2_{\mathrm {bare}}-1.5) \ ,
\no
Z_q^{\mathrm {RI'}}((2\,{\mathrm {GeV}})^2,g^2_{\mathrm {bare}}) 
&=& 0.805(14) - 0.323(6) \left(^{+0.002}_{- 0.040}\right)
\,(g^2_{\mathrm {bare}}-1.5) \ ,
\no
Z_q^{\overline{\mathrm {MS}}\,\mathrm{pert}}((2\,{\mathrm {GeV}})^2,
g^2_{\mathrm {bare}}) &=& 0.750(3)(7) - 0.313(20) 
\,(g^2_{\mathrm {bare}}-1.5) \ ,
\no
Z_q^{\overline{\mathrm {MS}}\,\mathrm{non-perturbative}}((2\,{\mathrm {GeV}})^2,
g^2_{\mathrm {bare}}) &=& 0.781(6)(21) - 0.313(20) \,(g^2_{\mathrm {bare}}-1.5) \ ;
\eeq 
where the systematic error is estimated from the scattering of the results 
in tables~\ref{tab:p2andA2}, \ref{tab:p2andA2_sliding}, \ref{tab:range} and
\ref{tab:merged}. We use the lattice spacings listed in table~\ref{setup}. 

Futhermore, table~\ref{tab:global_comp} also shows a nice agreement between the condensates
for $N_f=2$, although some systematics appears not to be under control for $N_f=0$. 
This supports the interpretation of the $\propto 1/p^2$ contribution as being 
due to a condensate of the only dimension two operator in Landau gauge: $A^2$.
Another confirmation comes from the validity of Martinelli-Sachrajda's
criterium~\cite{Martinelli:1996pk}. The accuracy on  $g^2 \VEV{A^2}$ is however limited due to several correlations
in the fits.  Further and more accurate checks of the consistency of $g^2
\VEV{A^2}$ from other renormalisation constants will be very welcome.


\section*{Acknowledgements} 
We thank Alain Le Yaouanc, Vittorio Lubicz and Gian-Carlo Rossi 
for a critical reading of this manuscript and the very useful subsequent discussions. 
This work was granted access to the HPC resources of CINES and IDRIS 
under the allocation 2010-052271 made by GENCI.  
J. R-Q is indebted to the Spanish MICINN for the 
support by the research project FPA2009-10773 and 
to ``Junta de Andalucia'' by P07FQM02962.
Z. Liu thanks the UK Science and Technology Facilities Council
(STFC) for financial support.


\appendix
\section{Appendix: The Wilson coefficients at $O(\alpha^4)$}
\label{appendix2}

The purpose of this appendix is to describe briefly the OPE analysis of the quark propagator 
renormalization constant defined in \eq{Zqdef} that leads us to
\eq{eq:zqfit}, where the four-loops 
results in ref.~\cite{Chetyrkin:2009kh} are exploited to derive the Wilson coefficients with 
the appropriate renormalization prescription. This OPE analysis 
is analogous to the one performed in refs.~\cite{Boucaud:2001st,Boucaud:2000nd,Boucaud:2005xn,Boucaud:2008gn}.
The starting point is the OPE of the inverse of the quark propagator:
\beq
S^{-1}(p^2,\mu^2) \ = \ Z^{-1}_q(\mu^2) \ S^{-1}_{\rm bare}(p^2)  &= &
\left(S^{\rm pert}\right)^{-1}(p^2,\mu^2) \ + \ i \pslash 
\frac{c_{2 Z_q}\left(\frac{p^ 2}{\mu^2},\alpha(\mu)\right)}{p^2} 
\frac{\langle A^2\rangle_{R,\mu^2}}{4 (N_c^2-1)} \delta_{ab}  
\nonumber \\
&=&
\frac{\left(S^{\rm pert}_{\rm bare}\right)^{-1}(p^2)}{ Z^{\rm pert}_q(\mu^2)}
\ + \ i \pslash 
\frac{c_{2 Z_q}\left(\frac{p^ 2}{\mu^2},\alpha(\mu)\right)}{p^2} 
\frac{\langle A^2\rangle_{R,\mu^2}}{4 (N_c^2-1)} \delta_{ab}  
\ , \nonumber \\
\label{ap-prop}
\eeq
where only the leading term in $\pslash$ is kept, the quark mass
being assumed to be negligible or to vanish. The cut-off
regularization dependence is omitted for the bare quantities but that  on the
renormalization momentum, $\mu$, is explicitly written for the renormalized
ones.  In the RI'-MOM scheme we define $Z^{\rm RI'}_q$ such that $(S_{\rm
bare})^{-1}(p^2)= i \pslash \delta_{ab}  \ Z^{\rm  RI'}_q(p^2)$  and $Z^{\rm
pert\; RI'}_q$ such that,  $(S^{\rm pert}_{\rm bare})^{-1}(p^2)= i \pslash
\delta_{ab} \ Z^{\rm pert \; RI'}_q(p^2)$. Then, the renormalization momentum,
$\mu^2$, taken to lie on the perturbative regime,  one can apply \eq{Zqdef} and
obtain, 
\beq
\frac{Z_q^{RI'}(p^2)}{ Z^{\rm pert}_q(\mu^2)} 
&=&
\frac{Z_q^{\rm pert\; RI'}(p^2)}{ Z^{\rm pert}_q(\mu^2)}
\ + \
c_{2 Z_q}\left(\frac{p^ 2}{\mu^2},\alpha(\mu)\right) \
\frac{\langle A^2\rangle_{R,\mu^2}}{4 (N_c^2-1) p^2}  
\nonumber \\
&=&
c_{0 Z_q}\left(\frac{p^ 2}{\mu^2},\alpha(\mu)\right)  \ + \
c_{2 Z_q}\left(\frac{p^ 2}{\mu^2},\alpha(\mu)\right) \
\frac{\langle A^2\rangle_{R,\mu^2}}{4 (N_c^2-1) p^2}  \ .
\label{ap-prop2}
\eeq
which implies a definition of $c_{0 Z_q}$ and where 
$c_{2Z_q}(p^2,\mu^2)$ is the Wilson coefficient of $g^2 \VEV{A^2}$. 
Although not yet specifying the renormalisation scheme of $Z^{\rm pert}_q(\mu^2)$, 
we know that
\beq
c_{0 Z_q}\left(1,\alpha(\mu)\right) &=& 1 + O(\alpha^2)  \ ,
\eeq
while $c_{2 Z_q}$ is known up to $O(\alpha^4)$ in the $\mMSB$ 
scheme~\cite{Chetyrkin:2009kh} and, in particular,  
$c_{2 Z_q}^{\mMSB}\left(1,\alpha(\mu)\right)$ is given in eq (18) of that paper
using $q^2=\mu^2$. Let us however keep in mind that 
\beq\label{trentedeux}
c_{2 Z_q}\left(1,\alpha(\mu)\right) &=& \frac{32 \pi} 3 \alpha(\mu)
\;\left(1+{\cal O}(\alpha(\mu))\right) = \frac
{8\,g^2(\mu)}{3}\;\left(1+{\cal O}(\alpha(\mu))\right) \ .
\eeq
%
Now, with the help of the appropriate renormalization constants, one can also
write \eq{ap-prop2} in terms of bare quantities:
\beq\label{ap-bar}
Z_q^{RI'}(p^2) &=& Z^{\rm pert }_q(\mu^2) \
 c_{0 Z_q}\left(\frac{p^2}{\mu^2},\alpha(\mu)\right)
\nonumber \\ 
  &+&   Z^{\rm pert}_q(\mu^2) Z_{A^2}^{-1}(\mu^2,\Lambda^2) 
 c_{2 Z_q} \left(\frac{p^2}{\mu^2},\alpha(\mu)\right) \
\frac{\langle A^2 \rangle}{4 (N_c^2-1) p^2} \ , 
 \eeq
where $A^2_R=Z^{-1}_{A^2} A^2$ . 
Then, as the $\mu$-dependence of both l.h.s. and r.h.s. of \eq{ap-bar} should
match  each other for any $p$, one can take the logarithmic derivative with
respect to $\mu$ and  infinite cut-off limit, term by term, on r.h.s. and
obtains: 
\beq\label{ap-diffeqs}
\gamma_q(\alpha(\mu)) + 
\left\{ \frac{\partial}{\partial\log\mu^2} 
+ \beta(\alpha(\mu))\frac{\partial}{\partial \alpha}\right\} \ 
\ln c_{0 Z_q}\left(\frac{q^2}{\mu^2},\alpha(\mu)\right) &=& 0 
\nonumber \\
-\gamma_{A^2}(\alpha(\mu)) + \gamma_q(\alpha(\mu)) 
+ \left\{ \frac{\partial}{\partial\log\mu^2} 
+ \beta(\alpha(\mu))\frac{\partial}{\partial \alpha}\right\} \ 
\ln c_{2 Z_q}\left(\frac{q^2}{\mu^2},\alpha(\mu)\right) &=& 0 \ ,
\eeq
where the $\beta$-function, chosen to be in $\mMSB$, is defined as
\beq\label{beta-f}
\beta(\alpha(\mu)) \ = \frac{d}{d\log\mu^2} \alpha(\mu) \ = \ - 4 \pi \ \sum_{i=0} \beta_i \left( \frac{\alpha(\mu)}{4 \pi}\right)^{i+2} 
\eeq
and where $\gamma_q(\alpha(\mu))$ and $\gamma_{A^2}(\alpha(\mu))$ are the anomalous
dimensions for the fermion propagator and local operator $A^2$, respectively, 
which are formally defined as
\beq
\gamma_{X}(\alpha(\mu)) \ = \ \frac{d}{d\log\mu^2} \log{Z_{X}} \ = -
\sum_{i=0} \gamma_i^{X} \ \left( \frac{\alpha(\mu)}{4\pi} \right)^{i+1} \ ,
\eeq
where $X$ stands for $q$ or $A^2$. 
The scheme for the anomalous dimension of $Z_{A^2}$ is imposed 
through the renormalization of the local operator $A^2$, as was done
in ref.~\cite{Boucaud:2001st} to obtain its leading logarithm contribution, and it is 
only known in the $\mMSB$ at the order ${\cal O}(\alpha^4)$ ~\cite{Gracey:2002yt}.
Then, that is the only possible choice of scheme for $\gamma_{A^2}$ 
in eqs.~(\ref{ap-diffeqs}). Concerning $Z_q^{\rm pert}$, its scheme is determined by the 
renormalization prescription for the non-perturbative propagator in the left hand-side 
of \eq{ap-prop}. Both $\mMSB$ and RI'-MOM are possible. As we aim to obtain 
a non-perturbative formula to be confronted to the lattice estimate of the RI'-MOM
quark renormalization constant, it is convenient also to prescribe the RI'-MOM scheme 
for  $Z_q^{\rm pert}$. Thus, eqs.~(\ref{ap-diffeqs}) must be re-written as
\beq\label{ap-diffeqs2}
\gamma_q^{\rm RI'}(\alpha(\mu)) + 
\left\{ \frac{\partial}{\partial\log\mu^2} 
+ \beta(\alpha(\mu))\frac{\partial}{\partial \alpha}\right\} \ 
\ln c_{0 Z_q}^{\rm RI'}\left(\frac{q^2}{\mu^2},\alpha(\mu)\right) &=& 0 
\nonumber \\
-\gamma_{A^2}^{\mMSB}(\alpha(\mu)) + \gamma_q^{\rm RI'}(\alpha(\mu)) 
+ \left\{ \frac{\partial}{\partial\log\mu^2} 
+ \beta(\alpha(\mu))\frac{\partial}{\partial \alpha}\right\} \ 
\ln c_{2 Z_q}^{\mWMSB}\left(\frac{q^2}{\mu^2},\alpha(\mu)\right) &=& 0 \ ,
\eeq
where 
\beq\label{eq:defc0}
c_{0 Z_q}^{\rm RI'} \left(\frac{p^ 2}{\mu^2},\alpha(\mu)\right)\equiv
\frac{Z_q^{\rm pert\; RI'}(p^2)}{Z^{\rm pert\; RI'}_q(\mu^2)} \ 
\eeq
and $c_{2 Z_q}^{\mWMSB}$ is in the $\mWMSB$ scheme\footnote{We define this scheme by imposing that the local operator of the Wilson expansion is renormalized in $\mMSB$, while the expanded operator (the quark propagator, in our case) is 
in a MOM scheme. We called this ``Wilson $\mMSB$". } explicitly defined by the second equation of (\ref{ap-diffeqs2}), 
after the RI'-MOM prescription for $Z_q^{\rm pert}$, that of $\MSB$ for $A^2$ and by the choice of a boundary condition, 
$c_{2 Z_q}^{\mWMSB}(1,\alpha(q))$. Then, from \eq{ap-prop2} one obtains
\beq\label{ap-fin0}
Z_q^{\rm RI'}\left(p^2\right) = Z^{\rm pert \; RI'}_q(\mu^2) \  
c_{0 Z_q}^{\rm RI'} \left(\frac{p^ 2}{\mu^2},\alpha(\mu)\right) \ 
\left( 1+ 
 \frac {c_{2 Z_q}^{\mWMSB} \left(\frac{p^2}{\mu^2},\alpha(\mu)\right)}
 { c_{0 Z_q}^{\rm RI'} \left(\frac{p^ 2}{\mu^2},\alpha(\mu)\right)} \ 
 \frac{\langle A^2 \rangle_{R,\mu^2}}{4 (N_c^2-1) p^2}\right) \ ,
\eeq
and, in practice, both eqs.~(\ref{ap-diffeqs2}) can
be combined to give the following differential equation,
\beq\label{ap-fin}
\left\{ -\gamma^{\mMSB}_{A^2}(\alpha(\mu)) +  \frac{\partial}{\partial\log\mu^2} 
+ \beta(\alpha(\mu))\frac{\partial}{\partial \alpha}\right\} \ 
\frac{c_{2 Z_q}^{\mWMSB}\left(\frac{p^2}{\mu^2},\alpha(\mu)\right)}
{c_{0 Z_q}^{\rm RI'}\left(\frac{p^2}{\mu^2},\alpha(\mu)\right)} \ = \ 0 \ ,
\eeq
that can be solved to provide us with the ratio of Wilson coefficients, $c_{2 Z_q}/c_{0 Z_q}$, 
required to implement \eq{ap-fin0}. For the purpose of the best comparison 
with the results from the analysis performed in ref.~\cite{Boucaud:2008gn}, we applied 
\beq\label{boundary}
c_{2 Z_q}^{\mWMSB}\left(1,\alpha(p)\right) \ \equiv \ c_{2 Z_q}^{\mMSB}\left(1,\alpha(p)\right)  \ ,
\eeq
where $c_{2 Z_q}^{\mMSB}\left(1,\alpha(\mu)\right)$ is taken from 
eq.~(18) of ref..~\cite{Chetyrkin:2009kh} using $q^2=\mu^2$, 
as a boundary condition which is equivalent to the one applied in the analysis of 
ref.~\cite{Boucaud:2008gn}.

On the other hand, if we take $Z_q^{\rm pert}$ to be renormalized in $\mMSB$, the 
equations in (\ref{ap-diffeqs}) reads 
\beq\label{ap-diffeqs3}
\gamma_q^{\mMSB}(\alpha(\mu)) + 
\left\{ \frac{\partial}{\partial\log\mu^2} 
+ \beta(\alpha(\mu))\frac{\partial}{\partial \alpha}\right\} \ 
\ln c_{0 Z_q}^{\mMSB}\left(\frac{q^2}{\mu^2},\alpha(\mu)\right) &=& 0 
\nonumber \\
-\gamma_{A^2}^{\mMSB}(\alpha(\mu)) + \gamma_q^{\mMSB}(\alpha(\mu)) 
+ \left\{ \frac{\partial}{\partial\log\mu^2} 
+ \beta(\alpha(\mu))\frac{\partial}{\partial \alpha}\right\} \ 
\ln c_{2 Z_q}^{\mMSB}\left(\frac{q^2}{\mu^2},\alpha(\mu)\right) &=& 0 \ ,
\eeq
where $c_{2 Z_q}^{\mMSB}$ is the Wilson coefficient computed in ref.~\cite{Chetyrkin:2009kh}, 
provided that the boundary condition, $c_{2 Z_q}^{\mMSB}\left(1,\alpha(\mu)\right)$, is taken again 
from eq.~(18) of the same paper using $q^2=\mu^2$. 
Then, we can combine again Eqs.~(\ref{ap-diffeqs3}) to obtain for $c_2^\mMSB/c_0^\mMSB$ the same 
equation \eq{ap-fin} that, with the same boundary condition, leads to:
\beq
\frac{c_{2 Z_q}^{\mWMSB}\left(\frac{p^2}{\mu^2},\alpha(\mu)\right)}
{c_{0 Z_q}^{\rm RI'}\left(\frac{p^2}{\mu^2},\alpha(\mu)\right)} 
\ = \
\frac{c_{2 Z_q}^{\mMSB}\left(\frac{p^2}{\mu^2},\alpha(\mu)\right)}
{c_{0 Z_q}^{\mMSB}\left(\frac{p^2}{\mu^2},\alpha(\mu)\right)} \ .
\eeq

On the other hand, we can also combine the second equation of (\ref{ap-diffeqs2}) with the second one of 
\eq{ap-diffeqs3} and obtain
\beq\label{ap-fin-alt}
\left\{ \gamma_q^{\rm RI'}(\alpha(\mu))  -\gamma_q^{\mMSB}(\alpha(\mu)) +  \frac{\partial}{\partial\log\mu^2} 
+ \beta(\alpha(\mu))\frac{\partial}{\partial \alpha}\right\} \ 
\frac{c_{2 Z_q}^{\mWMSB}\left(\frac{p^2}{\mu^2},\alpha(\mu)\right)}
{c_{2 Z_q}^{\mMSB}\left(\frac{p^2}{\mu^2},\alpha(\mu)\right)} \ = \ 0 \ ,
\eeq
that, according to \eq{boundary}, can be solved with the boundary condition 
$c_{2 Z_q}^{\mMSB}(1,\alpha(p))/c_{2 Z_q}^{\mMSB}(1,\alpha(p)) \equiv 1$ and leaves us with a relation of 
$\mWMSB$ and $\mMSB$ Wilson coefficients wich allows \eq{ap-fin0} to be re-written as
\beq\label{eq:app_final1}
Z_q^{\rm RI'}\left(p^2\right) &=& Z^{\rm pert \; RI'}_q(\mu^2) \  
c_{0 Z_q}^{\rm RI'} \left(\frac{p^ 2}{\mu^2},\alpha(\mu)\right) \ 
\nonumber \\
&\times& 
\left( 1+ 
 \frac {c_{2 Z_q}^{\mMSB} \left(\frac{p^2}{\mu^2},\alpha(\mu)\right)}
 { c_{0 Z_q}^{\rm RI'} \left(\frac{p^ 2}{\mu^2},\alpha(\mu)\right)} \ 
  \frac {c_{2 Z_q}^{\mWMSB} \left(\frac{p^2}{\mu^2},\alpha(\mu)\right)}
   { c_{2 Z_q}^{\mMSB} \left(\frac{p^ 2}{\mu^2},\alpha(\mu)\right)} \ 
 \frac{\langle A^2 \rangle_{\mMSB,\mu^2}}{4 (N_c^2-1) p^2}\right) \ .
\eeq
where, futhermore, $c_{2 Z_q}^{\mMSB}$ is to be taken from ref.~\cite{Chetyrkin:2009kh} and
$c_{0 Z_q}^{\rm RI'}$ from ref.~\cite{Chetyrkin:1999pq}. Thus, we can use either \eq{ap-fin0} with 
the solution of \eq{ap-diffeqs2} or \eq{eq:app_final1} with that of \eq{ap-fin-alt} to confront 
to the lattice estimates of $Z_q^{\rm RI'}$. Both expressions are equivalent. In the first case, 
one can proceed as was done in ref.~\cite{Blossier:2010ky} to solve \eq{ap-diffeqs2}. 
To illustrate this first method, let us remind that \eq{ap-diffeqs2} 
can be solved at the leading logarithm by applying the following ansatz,
\beq\label{ap-ansatz1}
\frac{\displaystyle c_{2 Z_q}^{\mWMSB}\left(\frac{p^2}{\mu^2},\alpha(\mu)\right)}
{\displaystyle c_{0 Z_q}^{\rm RI'}\left(\frac{p^2}{\mu^2},\alpha(\mu)\right)}
\ = \
\frac{32 \pi}{3} \alpha(p)
\ 
\left( \frac{\alpha(\mu)}{\alpha(p)}\right)^a \ 
\left( \rule[0cm]{0cm}{0.5cm} \ 1 + {\cal O}\left( \alpha \right) \ \right)
\ ,
\eeq
where we apply \eq{trentedeux} and the exponent $a$, required to satisfy \eq{ap-diffeqs2}, should be
\beq
a \ = \ \frac{\gamma_0^{A^2}}{\beta_0} \ = \frac{105-8 N_f}{132-8 N_f} \ .
\eeq
In the second case, to solve \eq{ap-fin-alt}, a similar ansatz extended to three-loops order 
can be applied, 
\beq\label{ap-ansatz2}
\frac{\displaystyle c_{2 Z_q}^{\mWMSB}\left(\frac{p^2}{\mu^2},\alpha(\mu)\right)}
{\displaystyle c_{2 Z_q}^{\mMSB}\left(\frac{p^2}{\mu^2},\alpha(\mu)\right)}
\ = \ 
\left( \frac{\alpha(\mu)}{\alpha(p)}\right)^b \ 
\left( \frac{\displaystyle 1+\sum_i r_i \ \left( \frac{\alpha(\mu)}{4\pi} \right)^i}
{\displaystyle 1+\sum_i r_i \ \left( \frac{\alpha(p)}{4\pi} \right)^i} \right) 
\ ,
\eeq
where we use \eq{boundary} for the boundary condition. Then, by requiring that the ansatz 
\eq{ap-ansatz2} verifies \eq{ap-fin-alt},  
the coefficients $b$ and $r_i$'s will be obtained in terms of those for the 
fermion propagator $\mMSB$ and RI'-MOM anomalous dimensions and for the $\mMSB$ 
$\beta$-function. 
However, in this case,
\beq
b \ = \ \frac{\gamma_0^{q \mMSB}-\gamma_0^{q {\rm RI'}}}{\beta_0} \ = \ 0 \ ,
\eeq
because the first-loop coefficient for the anomalous dimension is scheme independent 
(in the particular Landau gauge, this scheme-independent first-loop coefficient is also 
zero for any scheme~\cite{Chetyrkin:1999pq}). Furthermore, as can be seen in the appendix C 
of ref.~\cite{Chetyrkin:1999pq}, one is also left with $\gamma_1^{q \mMSB} \equiv \gamma_1^{q {\rm RI'}}$ 
in Landau gauge. Then, 
\beq\label{r1}
r_1 =\frac{\gamma_1^{q \mMSB}-\gamma_1^{q {\rm RI'}}}{\beta_0} \ = 0 \ , 
\eeq
and the Wilson coefficients for $\mMSB$ and RI'-MOM will thus differ only at 
the order ${\cal O}\left(\alpha^2\right)$, with the non-zero $r_i$'s coefficients 
to be applied in \eq{ap-ansatz2} given by
\beq\label{ris}
r_2 &=& \frac{ \gamma_2^{q \mMSB}-\gamma_2^{q {\rm RI'}} }{2 \beta_0} \ = 
- 25.4642 + 2.3333 \ N_f \ ,  \\ 
r_3 &=& \frac{\gamma_3^{q \mMSB}-\gamma_3^{q {\rm RI'}}}{3\beta_0} - 
\beta_1 \frac{\gamma_2^{q \mMSB}-\gamma_2^{q {\rm RI'}}}{3\beta_0^2} \ =
 \ -1489.9796 + 246.4424 \ N_f - 6.4609 \ N_f^2  \ ;
\nonumber 
\eeq
where the three and four-loop coefficient in $\mMSB$ and RI'-MOM for the fermion propagator anomalous 
dimension have been again obtained from ref.~\cite{Chetyrkin:1999pq}.
This leads, using Eqs.~(\ref{eq:app_final1},\ref{ap-ansatz2}--\ref{ris}) with
$N_c^2-1=8$ and $N_f=2$, to our final formulae for 
the free-of-artefacts lattice determination of $Z_q$ :
 \bea\label{eq:app_final2}
  Z_q^{\rm Latt\; artefree}(p^2,\beta)
   &=& Z_q^{\rm pert\; RI'}(\mu'^2) \,
  c_{0 Z_q}^{\rm RI'}\left(\frac{p^ 2}{\mu'^2},\alpha(\mu')\right)  \\
  &\times& \left( 1 \ + \,
  \frac{c_{2 Z_q}^{\mMSB}\left(\frac{p^ 2}{\mu^2},\alpha(\mu)\right)}
  {c_{0 Z_q}^{\rm RI'}\left(\frac{p^ 2}{\mu^2},\alpha(\mu)\right)} \
\frac{1 - 0.1317 \ \alpha^2(\mu) - 0.5155 \ \alpha^3(\mu) } {1 - 
0.1317 \ \alpha^2(p) - 0.5155 \ \alpha^3(p) } \
  \frac{\ \langle A^2\rangle_{\mMSB,\mu^2}}{32 \ p^2}  \right) \nonumber
  \eea
 In this last equation, we exploited the fact that the expression in parenthesis in 
 Eqs.~(\ref{eq:app_final1},\ref{eq:app_final2}) does not vary with the renormalization 
 momentum for the local operator $A^2$, as can be inferred 
 from~\eq{ap-fin}. Thus, once a given momentum, $\mu'^2$, is 
 fixed for the renormalization of the fermion propagator in the l.h.s. of 
 \eq{ap-prop}, the one appearing in $Z_q^{\rm pert \ RI'}$ in front of the expression 
 in parenthesis, one is still left with the the freedom of choosing a renormalization 
 momentum, $\mu^2$, which does not need to be the same, for the local 
 operator $A^2$ inside the parenthesis.  
 
 In \eq{eq:app_final2} the coefficients  $c_{0 Z_q}^{RI'}$ and $c_{2 Z_q}^{\mMSB}$ are known from 
 perturbation theory,  the former can be obtained from ref.~\cite{Chetyrkin:1999pq} and the 
 latter from ref.~\cite{Chetyrkin:2009kh}. Two parameters are to be fitted: 
  $Z_q^{\rm pert\; RI'}(\mu'^2)$ and the non perturbative condensate
 $g^2(\mu) \ \langle A^2\rangle_{R,\mu^2}$. It is important to underline
 that the condensate  {\it is defined via the OPE, i.e. from Eqs.~(\ref{eq:app_final1},\ref{eq:app_final2})}.
Its precise definition depends on the renormalisation scheme,the renormalisation scale, as well as the order in
 perturbation theory to which the coefficients $c_{0 Z_q}$ and $c_{2 Z_q}$
 are used. In \eq{eq:app_final1} and \eq{eq:app_final2} the renormalisation scheme 
 for $g^2(\mu) \ \langle A^2\rangle_{R,\mu^2}$ is $\mMSB$ and the
 scale is $\mu$ (10 GeV in our calculations). The coupling we use for the perturbative expansions 
of these coefficients, $c_{0 Z_q}$ and $c_{2 Z_q}$, is also chosen to be the $\mMSB$ one. 
These choices are kept all along the present paper. If we now wish to compare 
 $g^2(\mu) \ \langle A^2\rangle_{R,\mu^2}$ from the present calculation to 
 that from another calculation, for example from the strong coupling 
 constant~\cite{Blossier:2010ky}, {\it we must as far as possible use the same
 precise definition in both cases}. However, its dependence on the scheme and 
  on the order in perturbation theory is not so important: as seen in
 section~\ref{sec:running} other systematic uncertainties are larger.


\end{document}